\documentstyle[12pt,amsfonts]{article}

\newcommand{\del}{\nabla}

\newcommand{\p}{\partial}

\newcommand{\be}{\begin{enumerate}}
\newcommand{\ee}{\end{enumerate}}
\newcommand{\ba}{\begin{array}}
\newcommand{\ea}{\end{array}}
\newcommand{\beq}{\begin{equation}}
\newcommand{\eeq}{\end{equation}}
\newcommand{\bqa}{\begin{eqnarray}}
\newcommand{\eqa}{\end{eqnarray}}
\newcommand{\bqas}{\begin{eqnarray*}}
\newcommand{\eqas}{\end{eqnarray*}}

\begin{document}

\newtheorem{defi}{Definition}[section]
\newtheorem{lem}[defi]{Lemma}
\newtheorem{prop}[defi]{Proposition}
\newtheorem{theo}[defi]{Theorem}
\newtheorem{rem}[defi]{Remark}
\newtheorem{cor}[defi]{Corollary}

\newcommand{\qed}{\hfill $\Box$\vspace{.5cm}\medskip}


\title{Study of the family of Nonlinear Schr\"odinger equations by using the Adler-Kostant-Symes framework and the Tu methodology and their Non-holonomic deformation}

\author {Partha Guha\footnote{E-mail: {\tt partha@bose.res.in}}\\
S.N. Bose National Centre for Basic Sciences \\
JD Block, Sector III, Salt Lake \\ Kolkata - 700098,  India \\
\and
Indranil Mukherjee \footnote{E-mail: {\tt indranil.m11@gmail.com}}\\
Department of Natural Science \\ West Bengal University of Technology\\ BF 142, Salt Lake,
Kolkata-700064, India.\\
}

\date{today}

\maketitle



\abstract{The objective of this work is to explore the class of equations of the Non-linear Schrodinger type by employing the Adler-Kostant-Symes theorem and the Tu methodology.
In the first part of the work, the AKS theory is discussed in detail showing how to obtain the non-linear equations starting from a suitably chosen spectral problem.Equations
derived by this method include different members of the NLS family like the NLS, the coupled KdV type NLS, the generalized NLS, the vector NLS, the Derivative
NLS, the Chen-Lee-Liu and the Kundu-Eckhaus equations. In the second part of the paper, the steps in the Tu methodology that are used to formulate the hierarchy of
non-linear evolution equations starting from a spectral problem, are outlined. The AKNS, Kaup-Newell, and generalized DNLS hierarchies are obtained by using this
algorithm. Several reductions of the hierarchies are illustrated. The famous trace identity is then applied to obtain the Hamiltonian structure of these hierarchies
and establish their complete integrability. In the last part of the paper, the non-holonomic deformation of the class of integrable systems belonging to the NLS family is studied. Equations examined include the NLS, coupled KdV-type NLS and Derivative NLS (both Kaup-Newell and Chen-Lee-Liu equations). NHD is also applied to the hierarchy of equations in the AKNS system and the KN system obtained through application of the Tu methodology. Finally, we discuss the connection between the two formalisms and indicate the directions of our future endeavour
in this area }

\bigskip


\bigskip

{\bf Keywords and Keyphrases}. Adler-Kostant-Symes scheme,
Nonlinear Schr\"odinger equation, loop groups, bihamiltonian system,Tu methodology, Trace Identity, 
Non-holonomic deformation, Differential constraints.

\bigskip

\tableofcontents

\section{Introduction}

Completely integrable systems play an important role
in many physical applications including water waves,
plasma physics, field theory and nonlinear optics.
An important feature of many integrable evolution equations
is that a large class of their exact solutions,
particularly the solitons, can be derived by
applying the method of inverse scattering transform (IST)
in appropriate variables \cite{AbC,AKNS,FLT}. One of the most fascinating features of integrable
hierarchies is the fact that they possess a local bihamiltonian structure \cite{Magri};
these, in turn, yield the recursion operator and an infinite set of conserved
quantities. The bihamiltonian structure is a consequence of the existence
of classical $r$-matrices on the loop algebra.
The applications of Gelfand-Zakharevich \cite{GZ1,GZ2,GZ3}
bi-Hamiltonian structure, which is an extension of a
Poisson-Nijenhuis structure on phase space, has been extensively explored
by Falqui, Magri and Pedroni \cite{FMP1,FMP2,FMP3} in the context of separation of variables.
In \cite{Gu3} we unveil the connection between Adler-Kostant-Symes (AKS)
formalism applied to loop algebra and the Gelfand-Zakharevich
bi-Hamiltonian structure by superposition of the results of Fordy and Kulish \cite{FK}
in the AKS scheme. Fordy-Kulish decomposition has been demonstrated for the third-order flow
in \cite{Ath1}. Athorne and Fordy \cite{Ath2} generalized this to $(2+1)$-dimensions
and demonstrated how $N$-wave, Davey-Stewartson, and Kadomtsev-Petviashvili (KP) equations are
associated with homogeneous and symmetric spaces.
We have also shown \cite{Gu1,Gu2} that the AKS scheme also yields various $(1+1)$
dimensionl integrable equations which are various reductions of the SDYM equation.

\bigskip

It is well known that a systematic procedure of obtaining most finite
dimensional completely integrable systems is given by the
Adler, Kostant and Symes ( AKS) theorem \cite{Ad,AvM,Ko} applied to
 some Lie algebra ${\frak g}$ equipped with an
 ad-invariant non-degenerate bi-linear form. AKS scheme provides
a family of integrable systems, each consisting of a homogeneous space with a
hierarchy of flows generated by the $ad^{\ast}$-invariant functions.
We assume ${\frak g}$ be a vector space,
  presented as the linear sum of two subalgebras
 ${\frak g} = {\frak k} + {\frak  l}.$
 The bilinear form induces an isomorphism ${\frak g} \simeq {\frak g}^{\ast} $.
 Hence with the help of the bi-linear form
$< , >$ we can identify $ {\frak k}^{\ast} \sim {\frak l}^{\perp} $
and $ {\frak l}^{\ast} \sim {\frak k}^{\perp}$ where
\beq < {\frak k}^{\perp} , {\frak k} > = < {\frak l}^{\perp} , {\frak l} > = 0. \eeq
So ${\frak k}^{\perp}$ acquires a Poisson structure from that
 of ${\frak l}^{\ast}$. The co-adjoint action of $L$ on
 ${\frak k}^{\perp} \sim {\frak l}^{\ast}$ is given by
$$ g\circ p = \pi_{k^{\perp}}(gpg^{-1}) $$
for $g \in L $ and $ p \in {\frak k}^{\perp}$.
Then the infinitesimal action is
$$\eta (p) = \pi_{\frak k}^{\perp}[\eta , p]$$
for $\eta \in l $.

\bigskip

The symplectic manifold here is some co-adjoint $L$-orbit
 ${\cal M} \subset {\frak k}^{\perp} \simeq {\frak l}^{\ast} $.
 We associate to it a Hamiltonian equation of suitable ad-invariant
 function $f : {\frak g} \longrightarrow R $ for all $f|_{\cal M} $. Note that in this paper
our Lie algebra ${\frak g}$ is a loop algebra.\\

As we mentioned previously, many important equations can be derived from this approach,
 e.g. Adler and van Moerbeke \cite{Ad} obtained the Euler-Arnold equation as a
 geodesic flow on an ellipsoid, Ratiu \cite{Ratiu} obtained C. Neumann equation and so on.
These are all finite-dimensional systems. In order to use this technique to obtain partial
 differential equations, it is necessary to work with infinite-dimensional Lie algebras (IDLA).
This was demonstrated on loop algebras
by Reyman et al. \cite{RS1,RS2} and Flashchka et al.\cite{FNR}.
Hence AKS proves to be a very general  systematic procedure of obtaining many completely
integrable Hamiltonian system. \\

    It is also a well known fact that starting from a properly chosen spectral problem, one can set up a hierarchy of non-linear evolution equations. Obviously one of the most important challenges in the study of integrable systems is to find new such systems associated with non-linear evolution equations of physical significance. Another important issue in this context is to demonstrate the bi-Hamiltonian structure of the derived non-linear evolution equations which proves its complete integrability. When a set of non-linear evolution equations can be formulated as a Hamiltonian system in two distinct but compatible ways, then by a theorem due to Magri \cite{Magri}, they lead to an infinite sequence of conserved Hamiltonians that are in involution w.r.t either one of these two symplectic structures. One powerful approach for constructing infinite-dimensional Liouville integrable Hamiltonian systems is the one due to Tu \cite{Tu1,Tu2}. In this method, one uses the Trace Identity to derive the Hamiltonian structure of many integrable systems starting from an appropriate spectral problem. The related hierarchy of non-linear evolution equations can also be derived.\\

The motivation for the present work is to explore the family of Non-linear Schr\"odinger equations by using both the Adler-Kostant-Symes technique and the Tu methodology in parallel. The equations derived in the AKS framework are the Non-linear Schr\"odinger equation (NLSE), the coupled KdV type NLSE, the generalized NLSE, the vector NLSE, the Derivative NLSE, the Chen-Lee-Liu (CLL) type DNLS and the Kundu-Eckhaus equations.\\

The Tu methodology is used to first establish the AKNS hierarchy and then derive the NLS equations
and coupled KdV type NLS equations as special cases. The AKNS hierarchy was introduced in 1974 in \cite{AKNS}. In that work,
 the authors generalized the inverse scattering approach of Zakharov and Shabat that was developed in \cite{ZS} for a solution
of the NLS equation ( for details, see \cite{AbC}). The GNLSE is interpretated as a combination of the
ordinary NLS and the coupled KdV type NLS equations. The Tu formalism is next used to derive the hierarchy of
Kaup-Newell (KN) type non-linear evolution equations and in the lowest order the coupled KN system. After this,
the spectral problem is expanded and after imposing the appropriate constraint, the coupled Kundu type equation is
obtained. Under suitable reduction , this gives rise to the coupled KN, coupled CLL and the GI equations.
The multi-Hamiltonian structures of these systems of equations are examined using the trace identity.\\

An attempt is also made to rigorously examine the connection between the AKS formalism and the Tu methodology
and thus to unravel the relationship between these two powerful approaches to the construction and analysis of integrable systems.\\

At this point, we may mention that Non-holonomic Deformation (NHD) of integrable systems has come to occupy an important place in the literature on Integrable Systems. This is an interesting phenomenon in which an integrable system is perturbed in such a way that under suitable differential constraints on the perturbing function, the system maintains its integrability. In the last part of this paper, NHD of the family of NLS equations is studied including the NLS, coupled KdV type NLS, Derivative NLS (both Kaup-Newell and Chen-Lee-Liu systems). NHD is also applied to the hierarchy of equations in the AKNS system and the Kaup-Newell system obtained through the application of the Tu methodology.

\subsection{History of NLSE}
The Nonlinear Schr\"odinger (NLS) equation is a very well-known soliton equation \cite{AbC,FLT}. Various modifications and
generalizations of the NLS eqn have been considered. The four most important DNLS eqns are the Kaup-Newell (KN),
Chen-Lee-Liu (CLL), the GI and the Kundu eqns. The first generalization of the DNLS equation was considered
by Kaup and Newell \cite{KN}
\beq
iq_t = q_{xx} + i\beta (|q|^2q)_x,
\eeq
In our earlier work we showed that this equation and their generalizations are the
Euler-Poincar\'e flow on the space of first order scalar ( or matrix)
differential operators \cite{Gu2a}.

The immediate generalization of the Kaup-Newell equation was given by
Chen, Lee and Liu \cite{CLL}
\beq
iq_t = q_{xx} + i\alpha |q|^2q_x.
\eeq

The equivalence of Kaup-Newell \cite{KN} and Chen-Lee-Liu equation \cite{CLL}  was apparently
first noticed by Wadati and Sogo \cite{WS}, although it was believed by some mathematicians that
this was implicit in the work of Kaup and Newell.

\bigskip

 Gerdjikov and Ivanov \cite{GI} and independently Kundu \cite{Kundu1,Kundu2} proposed another version
of DNLS equation
\beq
iq_t = q_{xx} + i\beta q^2q_{x}^{\ast} + \frac{1}{2}\beta^2 q^3 {q^{\ast}}^3.
\eeq
It must be noted that Eckhaus \cite{CE,Eck} also derived this equation independently.
It is known that the KN, CLL and GI equations are described by using a
unified generalized derivative Schr\"odinger equation involving a parameter,
and their Hamiltonian structure and Lax pairs are also given
by unified and explicit formulae. Using suitable gauge transformations these equations
can be transformed into one another.

\bigskip

The method of gauge transformations can be applied to some more
generalized derivative nonlinear Schr\"odinger equation. Using this
process Kundu \cite{Kundu1,Kundu2} obtained
\beq
iq_t = q_{xx} + \beta |q|^2q + i\alpha (|q|^2q)_x,
\eeq
which is a hybrid of the NLS equation and Kaup-Newell system.
Actually this equation was proved to be integrable by Wadati, Konno and
Ichikawa {\footnote{ This was pointed out to me by late Professor Miki Wadati}}
and was further transformed by Kakei, Sasa and Satsuma \cite{KSS} into

\beq
iQ_T = Q_{XX} + 2i\gamma |Q|^2Q_X + 2i(\gamma - 1)Q^2Q_{X}^{\ast}
+ (\gamma - 1 )(\gamma - 2)|Q|^4Q,
\eeq
by means of change of variables
$$
q(x,t) = \sqrt{\frac{2}{\alpha}}Q(X,T)exp(i\frac{\beta}{\alpha}X +
i\frac{\beta^2}{\alpha^2}T),
$$
$$
x = X + \frac{2\beta}{\alpha}T, \qquad t = T, \qquad
\gamma = \frac{4\delta}{\alpha} + 2.
$$
Soliton solutions for $\gamma = 2$ were already known. Kakei et al. \cite{KSS}
have given the multi-soliton solutions for the general case.

\bigskip

Recently, while attempting to classify certain non-commutative
generalizations of classical integrable soliton equations,
Olver and Sokolov \cite{OS} made a detailed investigation
on the DNLS type systems of the form
\beq
P_t = P_{xx} + f(P,S,P_x, S_x) \qquad
S_t = - S_{xx} + g(P,S.P_x, S_x),
\eeq
where $P$ and $S$ take values in the associative algebra.
These two systems can be interpreted as nonabelian analogues
of the generalized derivative nonlinear Schr\"odinger equation.

In a recent paper, Tsuchida and Wadati \cite{TW} studied the Lax pair
of the matrix generalization of the Chen-Lee-Liu equation
\beq
iP_t = P_{xx} - iPSP_x, \qquad iS_t = -S_{xx} - iS_xPS,
\eeq
which is a member of the list by Olver and Sokolov \cite{OS}. In that paper they studied
the Lax pair of Chen-Lee-Liu equation.

\bigskip

There is also the higher order nonlinear Schr\"odinger
equation for the propagation of short light pulses in an optical
fibre. Theoretical prediction of Hasegawa and Tappert \cite{HT} that
an optical pulse in a dielectric fibre form an envelope soliton
and subsequent experimental verification by Mollenauer et al. \cite{MSG}
have made a significant impact in ultra high speed
telecommunications. This equation is given by
\beq
\p_z E = i(\alpha_1 \p_{tt} + \alpha_2 |E|^2E) + \alpha_3 \p_{ttt}E
+ \alpha_4 \p_t(|E|^2E) + \alpha_5 \p_t(|E|^2)E,
\eeq
where $E$ is the envelope of the electric field propagating in the $z$ direction
at a time $t$. The coefficients $\alpha_3$, $\alpha_4$, $\alpha_5$
respectively represent third order dispersion, self steepening
related to Kerr effect and the self frequency shifting via stimulated
Raman scattering\cite{ZS}. It is the last term which plays an important
role in the propogation of distortionless optical pulse over long distance.

\subsection{Background of the NHD formalism}

It was shown by Karasu-Kalkani et al [30] that the integrable 6th order KdV equation represented a Non-holonomic deformation (NHD) of the celebrated KdV equation
preserving its integrability band giving rise to an integrable hierarchy. The equation is given by
\beq
(\partial_x^2+8u_x \partial_x+4u_{xx})(u_t+u_{xx}+6u_x^2)=0
\eeq
With the change of variables $v=u_x$, $w=u_t+u_{xxx}+6u_x^2$, equation $(10)$ can be rewritten as a pair of equations,
\beq  \begin{array}{lc}
v_t+v_{xxx}+12vv_x-w_x=0 \\
w_{xxx}+8vw_x+4wv_x=0
\end{array}\eeq

The authors of [30] obtained the lax pair as well as an auto-B\"acklund transformation for equation (11). They claimed that equation (11) was
different from the KdV equation with the self consistent sources and wanted to explore the higher symmetries, higher conserved densities and Hamiltonian formalism for equation (11). In reference [45] Ramani et al bilinearized the KdV6 equation and deduced a new and simpler auto-B\"acklund transformation.

The terminology "nonholonomic deformation" was used by Kuperschmidt in [37]. Kuperschmidt rescaled $v$ and $t$ and modified equation (11) to take the following form
\beq \begin{array}{lc}
u_t-6uu_x-u_{xxx}+w_x=0 \\[.1cm]
w_{xxx}+4uw_x+2u_xw=0
\end{array}  \eeq

The pair of equations given by (12) can be converted into a bi-Hamiltonian system
\beq   \begin{array}{lc}
u_t=B_1 \left( \frac{\delta H_{n+1}}{\delta u}  \right)  - B_1(w) \\[.1cm]
\hspace{0.5cm} =B_2 \left( \frac{\delta H_n}{\delta u} \right)  - B_1(w),  \\
B_2(w)=0
\end{array}   \eeq

where the Hamiltonian operators are given by
\beq   \begin{array}{lc}
B_1\equiv \partial \equiv \partial_x   \hspace{2cm} \mbox{and}  \\[.1cm]
B_2\equiv \partial^3+2(u\partial + \partial u)
\end{array}   \eeq
 and $H_n$ denote the conserved densities.

 In reference [34], a matrix Lax pair, the N-soliton solution using the Inverse Scattering Transform (IST) technique as well as a two-fold integrable hierarchy were obtained by Kundu for the non-holonomic deformation of the KdV equation. The work was carried forward in reference [36] by Kundu et al to include the non-holonomic deformation of both KdV and mKdV equations as well as their symmetries, hierarchies and integrability. One of the authors of reference [36] extended the study to the NHD of the DNLS and the Lenells-Fokas equation in reference [35].  \\
Non-holonomic deformation of generalized KdV type equations were studied by Guha in reference [25] wherein a geometric insight was provided into the KdV6 equation. In this paper, Kirillov's theory of co-adjoint representation of the Virasoro algebra was used to generate a large class of KdV-6 type equations equivalent to the original equation. It was further shown that the Adler-Kostant-Symes approach provided a geometric formalism to obtain non-holonomic deformed integrable systems. NHD for the coupled KdV system was thereby generated. In reference [26], Guha extended Kupershmidt's infinite-dimensional construction to generate nonholonomic deformation of a wide class of coupled KdV systems, all of which follow from the Euler-Poincare-Suslov flows. In this paper, the author also derives a nonholonomic deformation of the N=1 supersymmetric KdV equation, also known as the sKdV6 equation. \\

\subsection{Organization}

We give a brief introduction to the Adler-Kostant-Symes (AKS) theory in Section 2.
We apply this scheme to current algebra over $S^1$ of a loop with a central extension given by
a two cocycle. The AKS scheme yields a hierarchy of commuting Hamiltonians.
We construct various types of nonlinear Schr\"odinger equations in Section 3, they
are associated to different symmetric spaces. Tu methodology is explained in Section 4.
Note that Tu's method is mainly confined to the Chinese group. In this paper we present
a description of this method and show how hierarchies of different integrable evolution equations can be constructed using this technique.
Section 5 is dedicated to the trace identity method. We derive the Hamiltonian structures for NLSE equations
using this method. Section 6 deals with the application of the NHD formalism to the class of equations belonging to the NLS family. We complete our work with a modest outlook in Section 7.

\section{Adler-Kostant-Symes Scheme}

Let $G$ be a connected compact semi-simple Lie group with the Lie
algebra ${\frak g}$, endowed with a non-degenerate
ad-invariant and symmetric inner product
$<.,.> : {\frak g} \times {\frak g} \longrightarrow
{\frak g}$, that is,
$$
<X,[Y,Z]> = <[X,Y],Z> \qquad \forall X,Y,Z \in {\frak g}.
$$
Its dual space ${\frak g}^{\ast}$ has a natural
Poisson structure
$$
 \{g_1 , g_2 \}(\mu) =
< \alpha , [ \frac{\delta g_1}{\delta \mu} ,
\frac{\delta g_2}{\delta \mu} ] >,
$$
of two smooth functions $g_1 $ and $g_2$ on ${\frak g}^{\ast}$.
The functional
derivative of $g$ ( or gradient of $g$) at $\mu$ is
the unique element $\frac{\delta f}{\delta \mu}$
of ${\frak g}$ defined by
$$
lim_{\epsilon \longrightarrow 0}\frac{1}{\epsilon}
[f(\mu + \epsilon \delta \mu) - f(\mu) ]
= < \delta \mu,\frac{\delta f}{\delta \mu} >.
$$

Here the gradient of $g_i$ are interpreted as elements of ${\frak g}$
due to the identification of ${\frak g} \simeq {\frak g}^{\ast \ast}$.

\bigskip

Let us introduce an additional structure from which,
in addition to the ordinary bracket,
a modified bracket can be defined as follows.

Let $$ R : {\frak g}  \longrightarrow {\frak g} $$
be an $R$-matrix, and this defines another Lie bracket on ${\frak g}$
\beq
 [ X , Y ]_{R} = \frac{1}{2} ([ RX , Y ] + [ X , RY ]),
\eeq
such a pair $({\frak g},R)$ is called a double Lie
algebra. It is known that $({\frak g},R)$ is a double Lie algebra
if and only if  the following bilinear map
$$
B_R : ({\frak g},R) \times ({\frak g},R) \longrightarrow ({\frak g},R)
$$
given by
\beq
B_R(X,Y) = [RX,RY] - R([X,Y]_R)
\eeq
is ad-invariant, that is, the equation
\beq
[X,B_R(Y,Z)] + [Y,B_R(Z,X)] + [Z,B_R(X,Y)] = 0
\eeq
holds for all $X,Y,Z \in {\frak g}$.

It is clear that the trivial solution $B_R(X,Y) = 0$
yields the Yang-Baxter equation.
The second solution  satisfies the so called
modified Yang-Baxter equation
\beq
B_R(X,Y) = -[X,Y].
\eeq

\bigskip

The best known class of $R$-matrices arises when the Lie algebra ${\frak g}$
split into a direct sum of two subalgebras
${\frak g} = {\frak g}_+ \oplus {\frak g}_-$.
Since there is a vector space decomposition of ${\frak g}$
into a direct sum of two Lie subalgebras,
hence, we put
\beq
R = P_+ - P_- ,
\eeq
where $P_{\pm}$ denotes the corresponding projection onto ${\frak g}_{\pm}$.
Under this identification the above bracket boils down to
$$ [X,Y]_R = [X_+ , Y_+] - [X_- , Y_-], $$
where $X_{\pm} = P_{\pm}X$.

\bigskip

By ${\frak g}^{\ast}$ and ${\frak g}_{R}^{\ast}$ we denote the
dual of ${\frak g}$ endowed with the Lie-Poisson structures arising from
$[.,.]$ and $[.,.]_{R}^{\ast}$ respectively. The Poisson bivectors
arising from the Lie brackets $[.,.]$ and $[,.,]_R$ are related by
$P_R = R^{\ast}P + PR$, where $R$ is considered to be a pointwise
lift of the map $R$ on ${\frak g}$ to the vector fields over ${\frak g}$
and $R^{\ast}$ is the transpose of this map.

The $R$-matrix construction
on ${\frak g}$ allows us to define  an additional  Lie-Poisson bracket
of the following form:
\beq
\{f,g\}(\mu) =  <\mu , [ R(\nabla f), (\nabla g)] + [ \nabla f , R(\nabla g)] >
\qquad f,g \in C^{\infty}({\frak g}^{\ast}).
\eeq

We wish to take a look at the special case of an $R$-structure given by
a splitting into two subalgebras. With ${\frak g} = {\frak g}_- \oplus
{\frak g}_+$, $R = P_+ - P_-$, $\mu \in {\frak g}^{\ast}$
 one computes the Lie-Poisson bracket arising from $[.,.]_R$:
\beq
\{f,g\}(\mu) =  2<\mu , [(\nabla f)_+, (\nabla g)_+]>
- 2<\mu, [(\nabla f)_- , (\nabla g)_-] >.
\eeq

\bigskip

\begin{defi}
We say a smooth function $ H : {\frak g}^{\ast}
\longrightarrow {\bf R} $
on any Lie algebra ${\frak g}$ is  $ Ad^{\ast} $ invariant  if
$$ H(Ad_{g}^{\ast}\alpha) = H(\alpha )$$
 for all $ \alpha \in {\frak g}^{\ast}$ and $ g\in G $.
\end{defi}

\bigskip

\begin{theo}(AKS)
Let $\frak g$ be Lie algebra with $R$-matrix $R : {\frak g}\longrightarrow
{\frak g}$, then the $ad^{\ast}$ invariant functions on ${\frak g}^{\ast}$
are in involution with respect to
$$
\{f,g\}(\mu) =  <\mu , [ R(\nabla f), (\nabla g)] + [ \nabla f , R(\nabla g)] >.
$$
Then the Hamiltonian flow on the coadjoint orbits in $L \subset \in
{\frak g}^{\ast}$ is
\beq
\frac{d}{dt}L = ad_{R(\nabla H)}^{\ast}L + R^{\ast}ad_{\nabla H}^{\ast}L,
\eeq
where $R^{\ast}$ is the transpose of $R$.
\end{theo}

\subsection{AKS theory and loop algebra}

The Adler-Kostant-Symes (AKS) theory produces hierarchies of completely
integrable partial (or ordinary) differential equations. This scheme is
quite general and is based on the following ingredients.

(a) A Lie algebra ${\frak g}$, with a non-degenerate bilinear form $<.,>$
which allows us to identify ${\frak g}$ with its dual ${\frak g}^{\ast}$. The
Lie algebra ${\frak g}$ splits into ${\frak g} = {\frak g}^+
\oplus {\frak g}^-$ i.e. two subalgebras ${\frak g}^+$ and ${\frak g}^-$.
The bilinear form is used to identify ${{\frak g}^{-}}^{\ast}$
with ${{\frak g}^{+}}^{\perp}$.

(b) The phase space is an $ad^{\ast}$ invariant finite dimensional
submanifold $\Gamma \subset {{\frak g}^{-}}^{\ast} \equiv
{{\frak g}^{+}}^{\perp}$. The Poisson structure on $\Gamma$ is the
Kostant-Kirillov structure associated to ${{\frak g}^{-}}^{\ast}$.

(c) The complete set of commutaing constants of motion will be
elements of the algebra $A(\Gamma)$ of ad-invariant functions
on ${\frak g}^{\ast}$ restricted to $\Gamma$

\subsection{Applications to loop group}

Let us apply this scheme to a loop group. Let $\Omega G$ be the space of based
loop, then the corresponding Lie algebra, called loop algebra, is the
Laurent polynomials in the variable $\lambda$ with coefficients in ${\frak g}$:
$$
\Omega {\frak g} = \{ X(\lambda ) = \sum_i x_i \lambda^i ; x_i \in {\frak g}\}, $$

with Lie bracket
$$
[X(\lambda),Y(\lambda)] := \sum_{i,j} [x_i,y_j]\lambda^{i+j}, \qquad
\hbox{  where }~ X(\lambda) = \sum x_i \lambda^i, \;
Y(\lambda) = \sum y_j \lambda^j.
$$

Here we can define the projection operator in the following way:

\[  P_{\pm}X = \left\{ \begin{array}{ll}
                   X & \mbox{if $ X = \sum_{n \geq 0} X_n \lambda ^n$ } \\
                  -X & \mbox{if $ X = \sum_{n < 0} X_n \lambda ^n$}
                    \end{array}
        \right. \]

We define the bilinear form on $\Omega {\frak g}$ as
$$
<X(\lambda),Y(\lambda) > := tr(\sum_{i+j = -1}x_iy_j)
= \oint tr(X(\lambda)Y(\lambda)) d\lambda.
$$

The two subalgebras of ${\Omega {\frak g}}$ are given as
$$
{\Omega {\frak g}}_+ := \{\sum_{0}^{k}g_i \lambda^i ~:~ g_i \in {\frak g} \},
\qquad {\Omega {\frak g}}_- :=
\{\sum_{- \infty}^{-1}g_i \lambda^i ~:~ g_i \in {\frak g} \}.
$$

With the above choice of inner product, one can verify
easily ${{\Omega {\frak g}}_-}^{\ast} = {{\Omega {\frak g}}_+}^{\perp}$,
so that $\Gamma$ can be identified
with a submanifold of ${{\Omega {\frak g}}_+}^{\perp}$:
$$
\Gamma := \{A(\lambda) = \sum_{0}^{n}a_{n-i}\lambda^i, \; n \hbox{ fixed } \}.
$$

\bigskip
The Kostant-Kirillov bracket for ${\widehat{ \Omega \frak g}}^{\ast}$
is given by
\beq
\{f,g\}(\mu) = <\mu, [\nabla f(\mu), \nabla g(\mu)] >, \qquad
\hbox{ where } \mu \in {\Omega \frak g}^{\ast}
\eeq

The gradient of a function $f~:~{\frak g}^{\ast} \longrightarrow {\bf C}$
is the vector field $\nabla f ~:~ {\frak g}^{\ast} \longrightarrow {\frak g}$
such that
$$
< \nabla f(\mu), X(\mu) > = df(X(\mu)) \qquad  \forall
\mu \in {\frak g}^{\ast}.
$$

But this does not restrict to ${\Omega \frak g}_{-}^{\ast}$. In fact,
with respect to this bracket, the Hamiltonian vector fields of elements
of $A(\Gamma)$ are identically zero, one justifies this by
$$ < ad_{X}^{\ast}\mu , \nabla H > = < \mu, [ X, \nabla H] > = 0$$ for
all $X \in  { \Omega \frak g}$.

\bigskip

Let us consider Hamiltonian equation with respect to $\{.,.\}$ where
$H$ can be expressed in terms of linear coordinates
$\mu_r = < \mu , X_r >$, where $X_r$ form the basis in ${ \Omega \frak g}_-$.
Thus the Hamiltonian equation becomes
$$
< \dot{\mu}, X_r > = \{H, \mu_r \}(\mu) =
< \mu, [(\nabla H(\mu))_-, (\nabla \mu_r(\mu))_-] >
$$
$$
\Longrightarrow \,\,\,\, < \dot{\mu}, X_r > \,=\, < \mu, [(\nabla H(\mu))_-, X_r] >
$$
$$
\Longrightarrow \,\,\,\, < \dot{\mu}, X_r > \,=\, < [(\nabla H(\mu))_+, \mu], X_r >,
$$
hence we obtain
\beq
\dot{\mu} = [(\nabla H(\mu))_+, \mu].
\eeq

\bigskip

\paragraph{Hierarchy equation} Let us consider the Hamiltonians
\beq
H_i(\mu) = \frac{1}{2}tr(\lambda^{-(p-i)}\mu^2, \qquad 0 \geq i \geq p.
\eeq
The gradient of $H$ is given by
$$
< \nabla H_i , X > = dH_i(X) = tr\,(\lambda^{-(p-i)}\mu X),
$$
hence $\nabla H_i = \lambda^{-(p-i)}\mu$.

Therefore, the Hamiltonian equations motion for $H_i$ are given as
\beq
\frac{d\mu}{dt_i} = [  (\lambda^{-(p-i)}\mu)_+, \mu ].
\eeq

\subsection{ AKS equation with cocycle}
Sometimes it is necessary to define orbit starting from the higher powers of
$\lambda$. In order to meet such demand we alter the Adler-Kostant-Symes
(AKS) scheme slightly. Instead of using the bilinear form in the previous section
we use
\beq
tr_n(X(\lambda)Y(\lambda)) := tr_0(\lambda^n X(\lambda)Y(\lambda)
\eeq

such that
$$
{{\Omega \frak g}_+}^{\perp} = \{X(\lambda) = \sum_{i \geq -n}^{m}x_i\lambda^i
\} \qquad {{\Omega \frak g}_-}^{\perp} =
\{Y(\lambda) = \sum_{i \leq -n-1}y_i\lambda^i \}.
$$

\bigskip

Hence the previous computation becomes
$$
< \dot{\mu}, X_r >_n = < [(\nabla H(\mu))_+, \mu], X_r >_n \qquad \hbox{ for }
\hbox{ all } X_r \in { \Omega \frak g}_-.
$$
So that again we have
$\dot{\mu} = [(\nabla H(\mu))_+, \mu]$.

\bigskip

Our first aim is to extend the loop algebra ${\Omega {\frak g}}$.
Let us introduce a non-trivial two cocycle on $\Omega {\frak g} $,
known as Maurer-Cartan cocycle.
Then corresponding to the centrally extended
Lie group $ \widehat{\Omega G} = {\Omega G} \times {\bf R} $
the Lie algebra is $ \widehat{\Omega {\frak g} } = \Omega {\frak g}
\oplus \bf R $. This is a centrally extended loop
algebra associated with $2$-cocycle $\omega(X,Y) = (X,\frac{dY}{dx})$.
Loop algebra $\widehat{ \Omega {\frak g}}$
satisfies the following commutation relation
$$ [(X,a),(Y,b)] = ([X,Y], \int_{S^{1}}tr(XY^{\prime}) ) $$
where $(X,a),(Y,b) \in \widehat{ \Omega {\frak g}} $.
We also define the bilinear form on $\widehat{\Omega {\frak g}}$ by
$$ <(X,a),(Y,b)> = ab + \int tr(XY). $$

\bigskip

In this case the ad-invariant function satisfies:

\begin{lem}
 Suppose $ H $ is $ Ad^{\ast}$ invariant function on ${\Omega \frak g}^{\ast}$
then
 $$ ad^{\ast}(\nabla H(\alpha),a)(\mu , 1) = ((ad^{\ast}(\nabla H(\mu))(\mu ) +
(\nabla H)^{\prime},0) $$
\end{lem}

\bigskip

The co-adjoint representation leaves invariant the hyperplanes
$  e = $ constant.
Note that from the above proposition and lemma we can conclude two things \\
(1)  The centre of the $\widehat{ \Omega \frak g}$ acts trivially on
$\widehat{\Omega \frak g}^{\ast}$, the space of $\widehat{\Omega \frak g}^{\ast}$ is a natural $G$-module.\\
(2) $\widehat{\Omega G} $ acts on $\widehat{\Omega \frak g}^{\ast}$
by a gauge transformation. \\

\bigskip

The Poisson bracket is
$$
\{f,g\}(\mu + c I) = < \hat{\mu}, [\hat{R}(\nabla \hat{f}), \nabla \hat{g} ]
+ [ \nabla f, \hat{R}(\nabla \hat{g}) ] >
$$
$$
= < \hat{\mu}, [\hat{R}(\nabla {f}), \nabla {g} ] +
\omega (R(\nabla f), \nabla g)I
+ [ \nabla f, {R}(\nabla {g}) ] + \omega (\nabla f, R(\nabla g))I >
$$
$$
= < \hat{\mu}, [\hat{R}(\nabla {f}), \nabla {g} ] + [\nabla f, {R}(\nabla {g})]
+ c < R\p (\nabla f), \nabla g > + c < \p \nabla f, R (\nabla g) >,
$$

where $\hat{R}$ is the $R$-matrix on $\hat{\frak g}$, it satisfies
 $$
\hat{R} : \tilde{\Omega \frak g} \longrightarrow \tilde{\Omega \frak g}
\qquad  \hat{R}(k + \alpha I) : = R(k).
$$

\bigskip

\begin{prop}
The Poisson bracket in the space of $\tilde{\Omega \frak g}^{\ast}$
for the two smooth functions has the form
$$ \{f_1 , f_2 \}(Y) = < [ \hat{R} (\nabla f_1) , \nabla f_2 ], Y >
+ [ \nabla f , R(\nabla g) ]
$$
$$
+ \int_{S^1} R \nabla f_1 \frac{d \nabla f_2 }{dx} +
\int_{S^1} {\nabla f_1} R\frac{d \nabla f_2 }{dx}, $$
where $ Y \in \Omega g $.
\end{prop}

\bigskip

If we repeat the previous steps we arrive at

\begin{theo}
The Hamiltonian equations of motion on the hyperplane of ${\widehat{
\Omega \frak g}}^{\ast}$ generated by the gradient of the Hamiltonian $H(L)$ ,
the ad-invariant function, have the form
\beq
 \frac{d\mu}{dt} = \frac{d(\nabla H(\mu))_-}{dx}
+ [(\nabla H)_- , \mu]_m \qquad \mu \in
{\widehat{ \Omega \frak g}}^{\ast}
\eeq
 so it denotes that the connection $ \mu dx + {\nabla H}dt $ on a cylinder
$ S^{1} \times {\bf R}$ is flat.
\end{theo}

\subsection{ Hermitian Symmetric Spaces and Integrable Systems}

Let $G$ be a semi-simple Lie group and ${\frak g}$ be the
corresponding Lie algebra. Let $M$ be a homogeneous space of
${\cal G}$, so, $M$ is a differentiable manifold on which $G$
acts transitively. There is a  homeomorphism of the
coset space $G/K$ onto $M$ for some isotropy subgroup
$K$ of $G$ at a point of $M$.
Let $k$ be the Lie algebra of $K$ and ${\frak g}$ satisfies
$$ {\frak g} = k \oplus m    \mbox{     }\hbox{  and  }  [k,k] \subset k $$
where $m$ is the vector space complement of $k$.
The Lie algebra ${\frak g}$ splits in such a way
that $M$ is equipped with two kinds of extra structure, these are :\\
(1) left translation of $m$ around $G$ gives
rise to a canonical connection on the principle $K$ bundle: ~ $G
\longrightarrow G/K $.\\
(2) When $ x \in M $, the map $ {\frak g} \longrightarrow T_x M $ given by
$$
\eta \longmapsto \frac{d}{dt}|_{t=0} exp (t\eta . x),
$$ restricts to give an isomorphism $[m]_x \longrightarrow T_x M $. \\
 The inverse map $$
\omega_x : T_x M  \longrightarrow [m]_x
$$
defines a ${\frak g}$-valued one form on $M$,
 known as Maurer Cartan form.

  If $k$ and $m$ satisfy
$$  [ k , m ] \subset m $$ then  $G/K$ is called
reductive homogeneous space [42]. We can associate to these spaces a canonical connection with curvature and torsion. Curvature and torsion at a fixed point $p$ are given purely in terms of the Lie bracket operation,
$$ ( R(X,Y)Z)_{p} = -[[X,Y]_k , Z]   \mbox{    }\hbox{   ,   }  X,Y,Z \in m $$
$$  T(X,Y)_{p} = -[X,Y]_m   \mbox{      }\hbox{   ,   }    X,Y \in m . $$
If $k$ and $m$ satisfy above two conditions and also satisfy
$$ [m,m] \subset k , $$
then $G/K$ is a symmetric space. Here the curvature satisfies
$$ (R(X,Y)Z)_{p} = -[[X,Y],Z]      \mbox{    }\hbox{  ,   } X,Y,Z \in m .$$
Here  $[X,Y] \in k $ happens automatically due to $[m,m] \in k $.

\bigskip

Let $h$ be the Cartan subalgebra of ${\frak g}$ which
is the maximal abelian subalgebra
of diagonalizable elements of ${\frak g}$.
In terms of the Weyl basis
$$ [ H_i , H_j ] = 0    \mbox{         }   \hbox{     ,       }
[ H_i , X_{\alpha} ] = \alpha (H_i) X_{\alpha} $$
$$ [ X_{\alpha} , X_{\beta} ] = N_{\alpha , \beta} X_{\alpha + \beta }                \mbox{         }    \hbox{          }  (\alpha + \beta  \in  \Delta ) $$
$$   = \sum_{i=1}^{mh} C_{\alpha , i} H_i    \mbox{     }   \hbox{       } (\alpha + \beta = 0 )   $$
   $$  = 0   \mbox{       }  \hbox{      }  ( \alpha + \beta  \not \in  \Delta  ,   \alpha + \beta  \neq  0 )  $$
For any $H_i \in  h $ and $ X_{\alpha}  \in  {\frak g}\ h $ and
$N_{\alpha , \beta}$ and $C_{\alpha , i} $ are structure constants and $\Delta$ is a set all roots.

\bigskip

The components $ R_{jkl}^{i} $ and $ T_{jk}^{i} $ of the curvature
and torsion with respect to a basis $X_i$ of $ T_p M $ are defined by
$$ R(X_k , X_l)X_j = R_{jkl}^{i}X_i  \mbox{   }\hbox{   ,  } T(X_j , X_k ) = T_{jk}^{i}X_i $$
and the component of the metric
$ g(X,Y) = tr\mbox{    }(ad(X)ad(Y)) $  is $ g_{ij} = g(X_i , X_j) $.

\bigskip

Let $\varrho$ be an element of $h$. We select the isotropy group $K$ such that its Lie algebra is $k$. This is given by the centralizer
$$ C_g (\varrho ) =  \{ X  \in  g   \hbox{       }  | [ X , \varrho ] = 0  \} $$
If $\varrho $  is regular i.e. the eigenvalues $\alpha (\varrho )$ of $ ad \hbox{    }\varrho $ are mutually distinct then $ C_g (\varrho ) = h$ and
here $ [h , m] \subset m $.

\smallskip

In this case since $ k = h $ hence the corresponding
coset space $G/K$ decomposition is essentially a Cartan decomposition.

\smallskip

When  $ k = C_g (\varrho )  \supset  h $,
then the eigenvalues $\alpha (\varrho)$ coalesce, and thus
$ C_g (\varrho)$ becomes larger than $h$. Hence in this case
the homogeneous space $G/K$ becomes smaller.

\bigskip

         In the case of Hermitian symmetric spaces $\alpha (\varrho )$ have eigenvalues $\{ 0,\pm \alpha \}$. Thus $g$ splits up into $$ g = k \oplus m^{+} \oplus m^{-} .$$
If we set $ X^0  \in  k $, $ X^{\pm } \in m^{\pm }$ for any $X \in g $.
$$ [ \varrho , X^0 ] = 0      \mbox{       } \hbox{      ,   }  [ \varrho , X^{\pm} ] =  \pm \alpha X^{\pm } $$ here eigenvalues $\alpha (\varrho )$ take the same eigenvalue for all $X^{\pm}  \in  m^{\pm}$. From the second commutation relation we can assert that hermitian symmetric space has almost complex structure.

\section{Examples of integrable systems}

In this section we derive several equations of the Nonlinear
Schr\"odinger family by using the Adler-Kostant-Symes (AKS) framework.

\subsection{Non-linear Schrodinger Equation}

In this case we consider the orbit
$$   L=\lambda A_1 + A_2   $$
			$$ A_1=  \left(\begin{array}{cc}
                                    \matrix -i & 0 \\
                                           0  & i  \\
                                    \end{array}\right)$$,

\beq A_2 = \left(\begin{array}{cc}
                                    \matrix 0 & q \\
                                           r  & 0  \\
                                    \end{array}\right),      \eeq     \\
which is basically the space part of the Lax pair.\\

To obtain the temporal part of the Lax pair we set
\beq \nabla{H}=\sum_{j=0}^{-\infty}{\lambda^{2+j}{h_{2+j}}} \eeq \\
The $h_i$ are obtained from the equation,
 \beq  [L,\nabla{ H} ] - (\nabla{ H})_x= 0 \eeq

Equating different powers of $\lambda$ and solving recursively, we obtain
 \beq  h_2=A_1 , h_1=A_2 , h_0=\left( \begin{array}{cc}
						\matrix  h_{03} &  \frac{i}{2}q_x \\
                                                -\frac{i}{2}r_x  &  -h_{03} \\
							\end{array} \right) \eeq
where $h_{03}$ is as yet undeterminate.\\
Now the AKS flow is given by,

\beq   L_t=(\Pi_n(\del{H}))_x - [L,\Pi_n(\del{H})]  \eeq

where,  $\Pi_n(\del{H})$ represents the projection of $\del{H}$ on the subalgebra containing non-negative powers of $\lambda$ i.e.
\beq    \Pi_n(\del{H})=\lambda^2 h_2+\lambda h_1+h_0    \eeq
Equating the terms containing $\lambda$ , we get $h_03=-\frac{i}{2}qr$ while the $\lambda$-free terms lead to the following dynamical equations,
\beq   \begin{array}{cc}
q_t=\frac{i}{2}q_{xx}-iq^2 r \\
r_t=-\frac{i}{2}r_{xx}+iqr^2
\end{array}    \eeq
which is the coupled NLS equations and reduce to the coventional NLS upon putting $r=q^*$.

\subsection{Coupled KdV type NLS equations}

We start with the same orbit viz that given by (29) but take
\beq \nabla{H}=\sum_{j=0}^{-\infty}{\lambda^{3+j}{h_{3+j}}} \eeq
and use it in equation (31). \\
Equating various powers of $\lambda$and solving recursively we get,

               \beq  h_3=A_1 , h_2=A_2 , h_1=\left( \begin{array}{cc}
					\matrix -\frac{i}{2}qr &  \frac{i}{2}q_x \\
                                          -\frac{i}{2}r_x  &  \frac{i}{2}qr \\
							\end{array} \right),
						 h_0=\left( \begin{array}{cc}
						\matrix  h_{03} &  \frac{q^2 r}{2}-\frac{1}{4}q_{xx} \\
                                                                           \frac{q r^2}{2}-\frac{1}{4}r_{xx}      &  -h_{03} \\
							\end{array} \right)
 \eeq
Next using the AKS flow equation (33), using  $\Pi_n(\del{H})$ and equating different powers of $\lambda$, we obtain
$h_{03}=\frac{1}{4}(r q_x - q r_x)$ which completes the determination of $h_0$.\\
The $\lambda$ indepedndent terms lead to the following equations:
\beq   \begin{array}{cc}
q_t=-\frac{1}{4}q_{xxx}+\frac{3}{2}q q_x r \\
r_t=-\frac{1}{4}r_{xxx}+ \frac{3}{2}r r_x q
\end{array}    \eeq
Putting $r=q^*$, we obtain the coupled KdV type NLS equations viz
\beq   \begin{array}{cc}
q_t=-\frac{1}{4}q_{xxx}+\frac{3}{2}|q|^2 q_x  \\
q_t^*=-\frac{1}{4}q_{xxx}^*+ \frac{3}{2}|q|^2 q_x^*
\end{array}    \eeq

\subsection{Generalized Nonlinear Schr\"odinger Equation}

Choose the orbit $L$ as before but now take  $$\del{H(\lambda)}=\del{H_1(\lambda)}+\xi \del{H_2(\lambda)}$$
where
\beq \nabla{H_1(\lambda)}=\sum_{j=0}^{-\infty}{\lambda^{2+j}{h_{2+j}}}  \hspace{4mm} and  \hspace{4mm} \nabla{H_2(\lambda)}=\sum_{j=0}^{-\infty}{\lambda^{3+j}{g_{3+j}}} \eeq
and use it in equation (31).
Now equating various powers of $\lambda$ we obtain the following:\\ \hspace{4mm}
$\lambda^4:$  \hspace{4mm}  $g_3=A_1$  \\  \hspace{4mm}
$\lambda^3:$  \hspace{4mm}  $[A_1,h_2]+\xi [A_1,g_2]+\xi [A_2,A_1] + \xi A_{1x}=0$  \\
which is satisfied on choosing,  \\
\hspace{4mm}  $g_2=A_2,\hspace{4mm}h_2=A_1$
By equating the other powers of  $\lambda$ , we obtain after some algebra,
$$ h_1=A_2, \hspace{4mm} g_1=\left( \begin{array}{cc}
						\matrix -\frac{i}{2}qr &  -\frac{i}{2}q_x \\
                                                                             \frac{i}{2}r_x  &  \frac{i}{2}qr \\
							\end{array} \right), \hspace{4mm}
				h_0=\left( \begin{array}{cc}
						\matrix -\frac{i}{2}qr &  -\frac{i}{2}q_x \\
                                                                             \frac{i}{2}r_x  &  \frac{i}{2}qr \\
							\end{array} \right),$$
		$$		g_0=\left( \begin{array}{cc}
						\matrix -\frac{1}{4}(q_x r-q r_x) &  \frac{1}{2}q^2 r-\frac{1}{4}q_{xx} \\
                                                                              \frac{1}{2}q r^2-\frac{1}{4}r_{xx}  &  \frac{1}{4}(q_x r-q r_x) \\
							\end{array} \right) $$
Taking projection on the subalgebra containing non-negative powers of $\lambda$, we obtain the coupled system of Generalized NLS equations as,
\beq \begin{array}{cc}
q_t=-\frac{i}{2} q_{xx}+iq^2 r+\xi (-\frac{1}{4}q_{xxx}+\frac{3}{2}q q_x r) \\
r_t=\frac{i}{2}r_{xx}-iqr^2+\xi (-\frac{1}{4}r_{xxx}+\frac{3}{2}q r r_{x})
\end{array}  \eeq

Putting $r=q^*$ leads us to
\beq
q_t=-\frac{i}{2}q_{xx}+i|q|^2q+\xi(-\frac{1}{4}q_{xxx}+\frac{3}{2}|q|^2q_x)
   \eeq

\subsection{Dimensionless Vector Nonlinear Schr\"odinger Equation (VNLSE), a Manakov system}

In this case we  choose a different orbit    $$L=\lambda A_1+A_2$$
where,

\beq    \begin{array}{cc}

A_1=\left( \begin{array}{ccc}
	\matrix{ i & 0 & 0 \\
                        0  & -i & 0 \\
                        0 & 0 & -i  \\}
		\end{array} \right),

\hspace{4mm}   A_2=\left( \begin{array}{ccc}
	\matrix{ 0 & -u^{(1)*} & -u^{(2)*} \\
                        u^{(1)}  & 0 & 0 \\
                        u^{(2)} & 0 & 0  \\}
		\end{array} \right),
\end{array}  \eeq

that is we move to a matrix representation in higher dimension.
Take     $ \nabla{H(\lambda)}=\sum_{j=0}^{-\infty}{\lambda^{2+j} h_{2+j}} $    and use it in  $(31)$   leading to

\beq  \begin{array}{cc}
 h_2=A_1,  \hspace{4mm}  h_1=A_2,  \hspace{4mm}
 h_0= \left( \begin{array}{ccc}
	\matrix{ h_0^{11}  &   \frac{i}{2} u_x^{(1)*}  &   \frac{i}{2} u_x^{(2)*}\\
                         \frac{i}{2} u_x^{(1)}  &   h_0^{22}   &    h_0^{23} \\
                       \frac{i}{2} u_x^{(2)}  &   h_0^{32}    &    h_0^{33}  \\}
		\end{array} \right)
\end{array}  \eeq

where entries in $h_0$ remain undetermined.   \\
Now using the AKS flow equation $(33)$ , after some lengthy calculations, we obtainc

\beq  \begin{array}{lc}
h_0^{11}=-\frac{i}{2}(u^{(1)*}u^{(1)}+u^{(2)*}u^{(2)})    \\
h_0^{22}=\frac{i}{2}  (u^{(1)}u^{(1)*})    \\
h_0^{23}=\frac{i}{2}  (u^{(1)}u^{(2)*})    \\
etc.
\end{array}       \eeq

and the dynamical systems

\beq  \begin{array}{lc}

iu_t^{(1)}+(|u^{(1)}|^2 + |u^{(2)}|^2)u^{(1)} + \frac{1}{2} u_{xx}^{(1)}=0    \\
iu_t^{(2)}+(|u^{(1)}|^2 + |u^{(2)}|^2)u^{(2)} + \frac{1}{2} u_{xx}^{(2)}=0

\end{array}       \eeq    \\ which are the desired equations.

\subsection{Derivative NLS equation (DNLS Eqn)}

We now begin to explore the different types of Derivative NLS equations using the AKS formalism.  \\
To obtain the DNLS eqn using the AKS technique we start with the orbit
\beq
L= \lambda^2 A_1+ \lambda A_2
\eeq
where  $A_1$  and  $A_2$  are defined previously.   \\
We choose  \beq    \nabla{H(\lambda)}=\sum_{j=0}^{-\infty}{\lambda^{4+j} h_{4+j}}      \eeq

Following the usual procedure, we obtain

\beq  \begin{array}{cc}
h_4=A_1,   \hspace{4mm} h_3=A_2,   \hspace{4mm}  h_2=\frac{1}{2} qrA_1,     \hspace{4mm}
 h_1= \left( \begin{array}{cc}
						\matrix   0   &  \frac{1}{2}q^2 r+\frac{i}{2}q_{x} \\
                                                                              \frac{1}{2}q r^2-\frac{i}{2}r_{x}  &    0
							\end{array} \right)
\end{array}   \eeq        \\

Next we impose the condition that  $\Pi_n (\del {H})$  should contain positive powers of  $ \lambda$  only, so that the  $h_0$  term gets dropped.    \\

Now using eqn  $(33)$,  we arrive at
\beq  \begin{array}{lc}

q_t=\frac{i}{2} q_{xx} + \frac{1}{2} (q^2 r)_x   \\
r_t=-\frac{i}{2} r_{xx} + \frac{1}{2} (q r^2)_x

\end{array}     \eeq    \\
On putting  $r=q^{*}$ , the above equations reduce to,
\beq  q_t=\frac{i}{2} q_{xx} + \frac{1}{2} (|q|^2 q)_x    \eeq   \\

This is the Kaup-Newell type DNLS eqn.

\subsection{Chen-Lee-Liu (CLL) type DNLS equation}

Here we take the orbit to be    $$ L=\lambda^2 A_1+\lambda A_2 + A_0$$    \\
where,
$$   \begin{array}{cc}
A_1= \left( \begin{array}{cc}
\matrix   -i   &  0 \\
               0  &    i
\end{array} \right), \hspace{4mm}
A_2= \left( \begin{array}{cc}
\matrix   0   &  q \\
               r  &    0
\end{array} \right),  \hspace{4mm}
A_0= \left( \begin{array}{cc}
\matrix   0   &  0 \\
               0  &    \frac{i}{2} qr
\end{array} \right)
\end{array}  $$   \\
and   \beq     \nabla{H}=\sum_{k=0}^{-\infty}{\lambda^{4+k} h_{4+k}}      \eeq   \\

The standard procedure outlined above leads to the following values of  $h_i$,    \\
$$  h_4=2A_1,   \hspace{4mm} h_3=2A_2,   \hspace{4mm}   h_2=qrA_1  \hspace{4mm}  \mbox{(by choice)} $$ \\
$h_1$  is chosen to be off-diagonal, and   \\
$$h_1= \left( \begin{array}{cc}
\matrix   0   &  i q_x+\frac{1}{2}q^2 r \\
      -i r_x+\frac{1}{2} q r^2  &    0
\end{array} \right)$$   \\
The off-diagonal elements of $h_1$ are determined by equating the powers of  $\lambda^3$.    \\
Equating powers of $\lambda^2$, leads to \hspace{4mm}    $[A_1,h_0]=0$  \\
whence, $h_0$  is determined to be
$$h_0= \left( \begin{array}{cc}
\matrix   0   &  0 \\
      0  &    -\frac{1}{2} (rq_x-r_x q)+\frac{i}{4} r^2 q^2
\end{array} \right)$$  \\
and thereby we obtain the CLL type DNLS eqn
\beq  \begin{array}{lc}
q_t=i q_{xx} + q q_x r   \\
r_t= -i r_{xx} + r r_x q   \\
\end{array}   \eeq

\subsection{Kundu-Eckhaus Equation}
Kundu \cite{Kundu1,Kundu2} and Eckhaus \cite{CE, Eck} independently derived what can now be called the Kundu–Eckhaus equation as a linearizable form of the nonlinear Schr\"odinger
equation.

\smallskip

The orbit $L$ and $\del{H}$ are taken to be the same as in $(52)$ but while taking the projection  $\Pi_n(\del{H})$ , only the terms containing the positive powers of $\lambda$ are considered i.e. we take
\beq  \Pi_n(\del{H})=\lambda^4 h_4 + \lambda^3 h_3 + \lambda^2 h_2 + \lambda h_1  \eeq    \\
Using $(31)$ and equating powers of $\lambda$, we obtain
$$A_{2t}=[h_1,A_0]+h_{1x}$$    \\
which leads to the Kundu DNLS equations:
\beq  \begin{array}{lc}
q_t=i q_{xx}+\frac{1}{2}q q_x r + \frac{1}{2} q^2 r_x + \frac{i}{4} q^3 r^2   \\
r_t= -i r_{xx} + \frac{1}{2} r r_x q + \frac{1}{2} r^2 q_x - \frac{1}{4} q^2 r^3   \\
\end{array}   \eeq

\section{Tu methodology}

The Tu method allows one to derive a hierarchy of non-linear evolution equations and also to obtain the Hamiltonian structure of these equations by using the trace identity. \\
Let us first focus on the method of obtaining the hierarchy of equations. To this end consider an isospectral problem of the form
\beq    \psi_x=U(\lambda)\psi    \eeq    \\
with $\lambda $ being the spectral parameter for which $\lambda_t=0$.     \\
Suppose that $U$ can be expressed in the form
\beq   U=R+u_1 e_1+u_2e_2+ .... +u_Pe_P    \eeq   \\
where the variables  $u_1,u_2, .... ,u_P\in S $  \hspace{4mm}  ($S$: Schwartz space)  \\
and $R,e_1,e_2, ...  ,e_P \in \Omega{g}$  \hspace{4mm}  ($\Omega{g}$: the loop algebra corresponding to the finite dimensional Lie algebra $g$)  and satisfy the conditions,  \vspace{2mm}  \\   \vspace{2mm}
(i)  $R,e_1,e_2, ....  ,e_P$ are linearly independent   \\
(ii) $R$ is pseudoregular.    \vspace{2mm}   \\
We first solve the stationary eqn,   \\
\beq     V_x=[U,V]     \eeq    for $V$, \\
and then search for a  $\triangle_n \in \Omega g$  such that for
\beq   V^{(n)}=(\lambda^n V)_+  + \triangle_n    \eeq    \\
it is found that
$$-V_x^{(n)} + [U,V^{(n)}] \in  e_1 +  e_2 + .... +  e_P  $$    \\
Once $\triangle_n$ is determined, the hierarchy of non-linear evolution eqns is determined from the zero-curvature representation,
\beq  U_t-V_x^{(n)} + [U,V^{(n)}]=0   \eeq

\subsection{The Ablowitz-Kaup-Newell-Segur (AKNS) hierarchy}

One starts with  $U=\left( \begin{array}{cc}
\matrix   -i\lambda   &  q \\
      r  &    i\lambda
\end{array} \right)$
and take \beq V=\left( \begin{array}{cc}
\matrix   a   &  b \\
      c  &    -a
\end{array} \right)  \eeq

Using these in $(58)$, we get the following relations: \beq \begin{array}{cc}
                                                        a_x=qc-rb \\
                                                        b_x=-2i\lambda b-2qa    \\
                                                        c_x=2i\lambda c+2ra
                                                        \end{array}  \eeq

Now expand $a,b,c$ as \\ $a=\sum_{m\geq 0} a_m \lambda^{-m}$ , $b=\sum_{m\geq 0} b_m \lambda^{-m}$ , $c=\sum_{m\geq 0} c_m \lambda^{-m}$ , \\
and put these in $(62)$, there by obtaining,
\beq \begin{array}{cc}
                                                        a_{mx}=qc_m-rb_m \\
                                                        b_{mx}=-2i b_{m+1}-2qa_m    \\
                                                        c_{mx}=2i c_{m+1}+2ra_m
                                                        \end{array}  \eeq
Choosing  $b_0=c_0=0$ and solving recursively we obtain the following values for $a_i,b_i$ and $c_i$.   \\   \hspace{4cm}
$a_0=\alpha$ (constant)
\beq \begin{array}{cc}
                                                        b_1=i\alpha q \\
                                                        c_1=i\alpha r    \\
                                                        a_1=0
                                                        \end{array}  \eeq

                                                        \beq \begin{array}{cc}
                                                        b_2=-\frac{\alpha}{2} q_x,  \hspace{4mm}
                                                        c_2=\frac{\alpha}{2} r_x,    \hspace{4mm}
                                                        a_2=\frac{\alpha}{2} qr
                                                        \end{array}  \eeq

                                                         \beq \begin{array}{cc}
                                                        b_3=\frac{i\alpha}{2} (-\frac{1}{2}q_{xx}+q^2 r)  \\
                                                        c_3=\frac{i\alpha}{2} (-\frac{1}{2}r_{xx}+q r^2)  \\
                                                        a_3=\frac{i\alpha}{4} (r q_x-q r_x)
                                                        \end{array}  \eeq

                                                         \beq \begin{array}{cc}
                                                        b_4=\frac{\alpha}{8} (q_{xxx}-6q q_x r)   \\
                                                        c_4=\frac{\alpha}{8} (-r_{xxx}+6q r r_x)   \\
                                                        \end{array}  \eeq  \\
                                                        and so on.   \\
Next we calculate the following expression  $$(\lambda^n V)_{+x}-[U,(\lambda^n V)_{+}]$$
Here $(\lambda^n V)_{+}$ denotes the terms in $(\lambda^n V)$ carrying non-negative powers of $\lambda$ only.
\beq (\lambda^n V)_{+}=\sum_{m=0}^{n} \lambda^{n-m}\left( \begin{array}{cc}
\matrix   a_m   &  b_m \\
      c_m  &    -a_m
\end{array} \right)  \eeq
Using $(61)$  and  $(68)$ we obtain,
$$(\lambda^n V)_{+x}-[U,(\lambda^n V)_{+}]=\sum_{m=0}^{n}\lambda^{n-m}\left( \begin{array}{cc}
\matrix{   a_{mx}-qc_m+rb_m   &  b_{mx}+2i\lambda b_m + 2q a_m \\
      c_{mx}-2i\lambda c_m - 2r a_m  &    -(a_{mx}-q c_m+r b_m)}
\end{array} \right)$$

On using the recurrence relations and simplifying the above matrix reduces to,
\beq  \left( \begin{array}{cc}
\matrix   0   &  -2ib_{n+1} \\
      2ic_{n+1}  &    0
\end{array} \right) \eeq

It is obvious that $\triangle_{n}=0$ and therefore we obtain the AKNS hierarchy,
\beq  \begin{array}{lc}
q_t=-2ib_{n+1}  \\
r_t=2ic_{n+1}
\end{array}  \eeq
Successive equations can be generated by putting $n=1,2,3 $ etc \\
Putting $n=2$, we obtain
\beq  \begin{array}{lc}
q_t=-2ib_{3} \\
r_t=2ic_{3}
\end{array}  \eeq
Using the values of $b_3$ and $c_3$ from $(66)$ we get
\beq  \begin{array}{lc}
q_t=\alpha(-\frac{1}{2} q_{xx}+q^2 r) \\
r_t=\alpha(\frac{1}{2} r_{xx}-q r^2)
\end{array}  \eeq
which constitute a system of NLS eqns.  \\
Further, setting $n=3$ leads to
\beq  \begin{array}{lc}
q_t=i\alpha (-\frac{1}{4} q_{xxx}+\frac{3}{2} q q_x r)  \\
r_t=i\alpha(-\frac{1}{4} r_{xxx}+\frac{3}{2} q r r_x)
\end{array}  \eeq
  which are a pair  of coupled KdV type NLS equation.

The Generalised Non-linear Schr\"odinder is essentially a combination of the ordinary NLS
and the coupled KdV type NLS equation and can be obtained similarly.

\subsection{The Derivative Non-linear Schr\"odinger Equation (Kaup-Newell hierarchy)}

Here we assume $U$ to have the form

\beq U= \left( \begin{array}{cc}
\matrix   -i\lambda^2   &  \lambda q \\
      \lambda r  &    i\lambda^2
\end{array} \right) \eeq
and use it in $(58)$ to solve for $V$ with $V$ chosen as
$$  \left( \begin{array}{cc}
\matrix   a   &  b \\
      c  &    -a
\end{array} \right) $$
Simple algebra yields
\beq  \begin{array}{lc}
a_x=\alpha (q c-r b)  \\
b_x=-2i\alpha^2 b - 2\lambda q a    \\
c_x=2i\lambda^2c+2\lambda ra
\end{array}  \eeq

Expanding the elements $a,b,c$ of the matrix $V$ as \\ $a=\sum_{m\geq 0} a_m \lambda^{-m}$ , $b=\sum_{m\geq 0} b_m \lambda^{-m}$ , $c=\sum_{m\geq 0} c_m \lambda^{-m}$ , \\ we arrive at the following recurrence relationships:

\beq  \begin{array}{lc}
a_{mx}=q c_{m+1} -r b_{m+1}  \\
b_{mx}=-2i b_{m+2} - 2 q a_{m+1}    \\
c_{mx}=2ic_{m+2} + 2r a_{m+1}
\end{array}  \eeq
Choosing $b_0=0,c_0=0,a_1=0$ we obtain
\beq  \begin{array}{lc}
a_0=\alpha ,b_1=2q,  c_1=2r     \\
b_2=0, c_2=0, a_2=-iq r    \\
b_3=iq_x+q^2r, c_3=-ir_x+qr^2, a_3=0   \\
b_4=0,  c_4=0,  a_4=\frac{1}{2}(r q_x-q r_x)-\frac{3}{2} r r_x q
\end{array}  \eeq
In general it is found that
\beq  \begin{array}{lc}
a_{2j+1}= 0 \\
b_{2j}=0    \\
c_{2j}=0
\end{array}  \eeq
for $j=0,1,2,...$,  hence let us write
\beq  \begin{array}{lc}
a=\sum_{j \geq 0} a_{2j} \lambda^{-2j} \\
b=\sum_{j \geq 0} b_{2j+1} \lambda^{-(2j+1)}   \\
c=\sum_{j \geq 0} c_{2j+1} \lambda^{-(2j+1)}
\end{array}  \eeq
so that

\beq V=\sum_{j \geq 0} \left( \begin{array}{cc}
\matrix   a_{2j} \lambda^{-2j}   &  b_{2j+1} \lambda^{-(2j+1)} \\
      c_{2j+1} \lambda^{-(2j+1)}  &    -a_{2j} \lambda^{-2j}
\end{array} \right)   \eeq

We now construct ${\tilde{V}}^{(n)}$ such that it contains positive powers of $\lambda$ only by defining
\beq
{\tilde{V}}^{(n)}=(\lambda^{2n+2} V)_{+} = \sum_{j=o}^{n}  \left( \begin{array}{cc}
\matrix    a_{2j} \lambda^{2(n-j)+2}   &  b_{2j+1} \lambda^{2(n-j)+1} \\
      c_{2j+1} \lambda^{2(n-j)+1}  &    -a_{2j} \lambda^{2(n-j)+2}
\end{array} \right)
\eeq

We now calculate
$${\tilde{V}}^{(n)}_{x}-[U,{\tilde{V}}^{(n)}]$$ and obtain,

\beq \begin{array}{lc}
e_{11}: \sum_{j=0}^{n} \lambda^{2(n-j)+2} [a_{2jx} - (q c_{2j+1} -r b_{2j+1})]    \\
e_{12}: \sum_{j=0}^{n} b_{(2j+1) x} \lambda^{2(n-j)+1} + 2\sum_{j=0}^{n} (i b_{2j+1} \lambda^{2(n-j)+3} + q a_{2j} \lambda^{2(n-j)+3} )  \\
e_{21}: \sum_{j=0}^{n} c_{(2j+1)x} \lambda^{2(n-j)+1} - 2\sum_{j=0}^{n} (i c_{2j+1} + r a_{2j}) \lambda^{2(n-j)+3}   \\
\end{array} \eeq

On using the recurrence relations, the diagonal element vanishes while the off-diagonal elements yield $\lambda b_{(2n+1)x}$ and $\lambda c_{(2n+1)x}$ respectively. \\
Thus
\beq
{\tilde{V}}^{(n)}_{x}-[U,{\tilde{V}}^{(n)}]=\lambda \left( \begin{array}{cc}
\matrix   0   &  b_{(2n+1)x} \\
      c_{(2n+1)x}  &    0
\end{array} \right)
\eeq
from the above it is clear that $\triangle_n=0$ whence, ${\tilde{V}}^{(n)}$ becomes equal to $V^{(n)}=(\lambda^{2n+2} V)_{+}$
The AKS equation now yields
\beq  \begin{array}{lc}
q_t=b_{(2n+1)x}   \\
r_t=c_{(2n+1)x}
\end{array}  \eeq
which is the hierarchy of non-linear evolution equations for Kaup-Newell system.
Putting $n=1$, we obtain
\beq  \begin{array}{lc}
q_{t_{1}} = b_{3x} =i q_{xx} + q^2 r_x + 2q q_x r    \\  \vspace{4mm}
r_{t_{1}} = c_{3x} =-i r_{xx} + r^2 q_x + 2 r r_x q
\end{array}   \eeq

Setting $n=2$ gives a higher order eqn of the system,
\beq  \begin{array}{lc}
q_{t_{2}} = b_{5x} = -\frac{1}{2} q_{xxx} + \frac{3}{4} (q^3 r^2)_x + \frac{3i}{2} (q q_x r)_x    \\    \vspace{4mm}
r_{t_{2}} = c_{5x} = -\frac{1}{2} r_{xxx} + \frac{3}{4} (q^2 r^3)_x - \frac{3i}{2} (r r_x q)_x
\end{array} \eeq

\subsection{The Derivative Non-linear Schr\"odinger equations - a general structure}
Let us expand the orbit by including another field variable $s$ in the diagonal term, whence $U$ takes the form

\beq
\left( \begin{array}{cc}
\matrix   -i\lambda^2-is   &  \lambda q \\
      \lambda r  &    i\lambda^2+is
\end{array} \right)
\eeq
as before, taking
$ V=
\left( \begin{array}{cc}
\matrix   a   &  b \\
      c  &    -a
\end{array} \right)
$
and expanding the elements $a$, $b$ and $c$ we obtain the recurrence relations
\beq \begin{array}{cc}
a_{mx}=q c_{m+1}-r b_{m+1} \\
b_{mx}=-2i b_{m+2}-2i s b_m -2q a_{m+1}    \\
c_{mx}=2i c_{m+2} + 2i s c_m + 2ra_{m+1}
\end{array}  \eeq
from which it can be shown that
\beq
a_{(m+1)x}=(r s b_m - q s c_m) - \frac{i}{2} (q c_{mx} + r b_{mx})
\eeq
Solving the system $(88)$ we obtain successively,
\beq  \begin{array}{lc}
a_0=-2i, b_0=0, c_0=0   \\
a_1=0, b_1=2q, c_1=2r   \\
a_2=-iq r, b_2=0, c_2=0   \\
a_3=0, b_3=i q_x - 2 q s + q^2 r, c_3=-i r_x - 2 r s + q r^2   \\
a_4=\frac{1}{2} (r q_x - q r_x) - \frac{3i}{4} q^2 r^2 + 2 i s q r, b_4=0, c_4=0
\end{array}  \eeq
and so on  \\
In general then it is found that
\beq
a_{2j+1}=0, b_{2j}=0, c_{2j}=0
\eeq
for $j=0,1,2,3,....$

\beq V=\sum_{j \geq 0} \left( \begin{array}{cc}
\matrix   a_{2j} \lambda^{-2j}   &  b_{2j+1} \lambda^{-(2j+1)} \\
      c_{2j+1} \lambda^{-(2j+1)}  &    -a_{2j} \lambda^{-2j}
\end{array} \right)   \eeq
As before taking ${\tilde{V}}^{(n)} = (\lambda^{2n+2} V)_{+}$ and evaluating the expression ${\tilde{V}}^{(n)}_x -[U,{\tilde{V}}^{(n)}]$
we obtain
\beq
{\tilde{V}}^{(n)}_x -[U,{\tilde{V}}^{(n)}]=\lambda (b_{(2n+1) x} + 2is b_{2n+1}) e_{12} + \lambda (c_{(2n+1) x} + 2is c_{2n+1}) e_{21}
\eeq
Since there is no diagonal element on the RHS of the above, we define
\beq
V^{(n)}={\tilde{V}}^{(n)} + \triangle_{n}
\eeq
where $\triangle_{n}$ is taken to be of the form
\beq
\triangle_{n} = \left(  \begin{array}{cc}
\delta_{n}  &  0  \\
0  &  -\delta_{n}
\end{array}
\right)
\eeq
Using equations  $(92) - (94)$ in the zero curvature equation  $(60)$ , we are led to the following dynamical equations.
\beq  \begin{array}{l}
-is_t = \delta_{nx}  \\
q_t = b_{(2n+1) x} +2is b_{(2n+1)} +2q \delta_{n}   \\
r_t = c_{(2n+1) x} - 2is c_{(2n+1)} -2r \delta_{n}
\end{array}
\eeq
But $\delta_{n}$ is yet undetermined.
To determine $\delta_n$, let us impose the condition
\beq
s=\beta q r
\eeq
where $\beta$ is a constant.
Using $(97)$ in the system of equations $(96)$ we obtain after simplification,

\beq   \begin{array}{l}
\delta_n= 2\beta \partial^{-1} [(r s b_{2n+1} - q s c_{2n+1})  - \frac{i}{2} (q c_{(2n+1) x} + r b_{(2n+1) x})]   \\
= 2\beta \partial^{-1} a_{[2(n+1) x]}    \\
=2 \beta a_{2(n+1)}
\end{array}   \eeq
where we have used  $(89)$  to simplify the expression in the square bracket.    \\
Putting back  $(98)$  in  $(96)$  we obtain the following dynamical systems
\beq  \begin{array}{l}
q_t= b_{(2n+1) x} + 2i \beta q r b_{2n+1} + 4 \beta q a_{2(n+1)}   \\
r_t = c_{(2n+1) x} - 2i \beta q r c_{2n+1} - 4 \beta r a_{2(n+1)}
\end{array}  \eeq
These represent a coupled system of hierarchy of equations.  \\
Putting  $n=1$, we obtain,
\beq  \begin{array}{l}
q_t = b_{3x} + 2i \beta q r b_3 + 4 \beta q a_4   \\
r_t = c_{3x}  - 2i \beta q r c_3 - 4 \beta r a_4
\end{array}  \eeq
After using  $s = \beta q r$  , we obtain the following expressions for  $b_3, c_3$  and  $a_4$  viz.
\beq   \begin{array}{l}
b_3= i q_x - (2 \beta -1) q^2 r   \\
c_3= -i r_x - (2 \beta -1) q r^2   \\
a_4= \frac{1}{2} (r q_x - q r_x) + (2 \beta - \frac{3}{4})i q^2 r^2
\end{array}  \eeq
Hence  $ (100)$  yields,
\beq  \begin{array}{l}
q_t= i q_{xx} - (4 \beta -1) q^2 r_x -2(2 \beta -1) q q_x r +i \beta (4 \beta -1) q^3 r^2   \\
r_t= -i r_{xx} - (4 \beta -1) r^2 q_x -2 (2 \beta -1) r r_x q -i \beta (4 \beta -1) q^2 r^3
\end{array}   \eeq
Eqns  $(102)$ represent coupled Kundu type systems.   \\
Several reductions of  $(102)$  are possible.   \\
Putting  $\beta = 0$, we get,
\beq   \begin{array}{l}
q_t = i q_{xx} + (q^2 r)_x    \\
r_t = -i r_{xx} + (q r^2)_x
\end{array}   \eeq
which form a coupled Kaup-Newell (KN) system.  \\
$ \beta = \frac{1}{4}$ leads to,
\beq   \begin{array}{l}
q_t = i q_{xx} + q q_x r    \\
r_t = -i r_{xx} + r r_x q
\end{array}   \eeq
which is the coupled Chen-Lee-Liu (CLL) system.   \\
Finally,  $ \beta = \frac{1}{2}$  yields,
\beq   \begin{array}{l}
q_t = i q_{xx} - q^2 r_x + \frac{i \beta}{2} q^3 r^2      \\
r_t = -i r_{xx} - r^2 q_x - \frac{i \beta}{2} q^2 r^3
\end{array}   \eeq
a coupled GI system.   \\
Putting  $ r = q^*$ in the above system of equations leads to further reductions.    \\
eg. setting  $ r = q^*$  in  $(97)$  leads to
\beq
q_t = i q_{xx} - (4 \beta -1) q^2 q_x^* - 2 (2 \beta -1) |q|^2 q_x + i \beta (4 \beta -1) |q|^4 q
\eeq
which represents Kundu type equation.

\section{The trace identity}

The trace identity is a powerful tool for constructing infinite-dimensional Liouville integrable Hamiltonian systems.
Starting from a properly chosen spectral problem, many integrable hierarchies and their Hamiltonian structures can be obtained by using trace identity method.   \\
As noted before, let $g$ be a finite dimensional semi-simple Lie algebra and $\Omega g$ the corresponding loop algebra.   \\
The Killing-Cartan  form $<x,y>$ is taken to be equal to $tr(xy)$ where $x,y\in g$ i.e. $<x,y>=tr(xy)$  \\
Let $U=U(\lambda, u)$ be an element of $\Omega g$ that depends on $\lambda$ and $u=(u_i)$, where $u_i$ are the field variables.   \\
For any solution $V$ of the stationary eqn $(58)$, which is of homogeneous rank, there exists a constant $\gamma$ such that for $\bar{V}=\lambda^{\gamma} V$ which is again a solution of $(58)$ it is true that
\beq  \begin{array}{lc}
\frac{\delta}{\delta u_i} <\bar{V},\frac{\partial U}{\partial \lambda}>=\frac{\partial}{\partial \lambda}<\bar{V}, \frac{\partial U}{\partial u_i}>
\end{array}  \eeq

This is the famous trace identity which on putting $\bar{V}=\lambda^{\gamma} V$ reduces to
\beq  \begin{array}{lc}
\frac{\delta}{\delta u_i} <V,\frac{\partial U}{\partial \lambda}>=(\lambda^{-\gamma} \frac{\partial}{\partial \lambda} \lambda^{\gamma}) <V, \frac{\partial U}{\partial u_i}>
\end{array}  \eeq

Here $\frac{\delta}{\delta u_i}$ stands for the variational derivative given by

\beq  \begin{array}{lc}
\frac{\delta F}{\delta u}=\frac{\partial F}{\partial u} - (\frac{\partial F}{\partial u_x})_x + (\frac{\partial F}{\partial u_{xx}})_{xx} - .....
\end{array}  \eeq

Further let $J$ and $L$ be two linear operators mapping $S^M$ into itself. Here $S$ denotes the Schwartz space over $\Re$ and $S^M=S\otimes ..... \otimes S$ ($M$ times)
where $\otimes$ denotes the direct/outer/tensor product.   \\

Suppose that  \vspace{2mm}    \\
(i) both $J$ and $JL$ are skew symmetric i.e.
\beq  \begin{array}{lc}
J^*=-J   \hspace{4mm}  \mbox{and} \hspace{4mm}  JL=L^*J
\end{array}   \eeq
(ii) there exists a series of scalar functions $\{H_n\}$, for which it is true that
\beq  \begin{array}{lc}
L^nf(u)=\frac{\delta H_n}{\delta u}
\end{array}  \eeq
for some $f(u)\in S^M$.   \\
Then $\{H_n\}$ is a common series of conserved derivatives for the whole hierarchy of equations.
\beq   \begin{array}{lc}
u_t=J L^n f(u)
\end{array}   \eeq
and further
\beq   \begin{array}{lc}
\{H_n, H_m\}=0
\end{array}   \eeq
 The above conditions are used to investigate the Hamiltonian structures of several integrable systems in the following sections.

\subsection{The Hamiltonian structure of the AKNS hierarchy}

 In this case, $U$ and $V$ are given as in $(61)$ and the hierarchy is given by $(70)$.   \\
 The hierarchy $(70)$ can be rewritten as
 \beq   \begin{array}{cc}
 u_t=\left(\matrix{q  \\  r}\right)_t=\left( \matrix{ -2i b_{n+1}  \\  2i c_{n+1}}\right)=J \left( \matrix{ i c_{n+1}  \\  i b_{n+1}} \right)
 \end{array}   \eeq
 where
 \beq   \begin{array}{lc}
 J=\left( \matrix{ 0  &  -2  \\  2  &  0} \right)
 \end{array}  \eeq
 Further the operator
 \beq   \begin{array}{lc}
 L= \left( \matrix{ -\frac{i}{2} \partial + i r \partial^{-1} q  &  -i r \partial^{-1} r  \\  iq \partial^{-1} q  &  \frac{i}{2} \partial - i q \partial^{-1} r} \right)
 \end{array}   \eeq
 is such that
 \beq   \begin{array}{lc}
 L^n \left( \matrix{c_1  \\  b_1} \right)=\left( \matrix{c_{n+1}  \\  b_{n+1}} \right)
 \end{array}   \eeq
 To check this we note that
 $$ \begin{array}{lc}  L \left( \matrix{c_1  \\  b_1} \right)= \left( \matrix{ -\frac{i}{2} \partial + ir \partial^{-1} q  &  -i r \partial^{-1} r  \\  i q \partial^{-1} q
 &  \frac{i}{2} \partial - i q \partial^{-1} r} \right) \left( \matrix{i \alpha r  \\  i \alpha q} \right)=  \left( \matrix{ \frac{\alpha}{2} r_x  \\  -\frac{\alpha}{2} q_x} \right)
 =\left( \matrix{c_2  \\  b_2} \right)
 \end{array}$$
 and
 $$  \begin{array}{lc}
 L^2 \left( \matrix{c_1  \\  b_1} \right)= \left( \matrix{ -\frac{i}{2} \partial + ir \partial^{-1} q  &  -i r \partial^{-1} r  \\  i q \partial^{-1} q
 &  \frac{i}{2} \partial - i q \partial^{-1} r} \right) \left( \matrix{\frac{\alpha}{2} r_x  \\  -\frac{\alpha}{2} q_x} \right)= \left( \matrix{-\frac{i \alpha}{4} r_{xx} + \frac{i \alpha}{2} q r^2
 \\  -\frac{i \alpha}{4} q_{xx} + \frac{i \alpha}{2} q^2 r} \right)= \left( \matrix{c_3  \\  b_3} \right)
 \end{array}  $$
 which proves assertion $(117)$ in these special cases.
 With
 \beq   \begin{array}{lc}
 J=\left( \matrix{0  &  -2  \\  2  &  0} \right), \hspace{4mm}  \mbox{we have}   \hspace{4mm}    J^*=\left( \matrix{0  &  2  \\  -2  &  0} \right)
 \end{array}   \eeq
 Thus,  $$J^*=-J$$
 Further,
 \beq   \begin{array}{lc}
 L^*=\left( \matrix{\frac{i}{2} \partial - i q \partial^{-1} r  &  -i q \partial^{-1} q  \\  i r \partial^{-1} r  &  -\frac{i}{2} \partial + i r \partial^{-1 q}} \right)
 \end{array}   \eeq
 We also note that
 $$   \begin{array}{lc}
 JL=L^*J=\left( \matrix{-2 i q \partial^{-1} q  &  -2(\frac{i}{2} \partial - i q \partial^{-1} r)  \\  2(-\frac{i}{2} \partial + i r \partial^{-1} q)  &  -2 i r \partial^{-1} r} \right)
 \end{array}   $$
 For using the trace identity we compute the following,
 \beq   \begin{array}{lc}
 \left< V, \frac{\partial U}{\partial \lambda} \right> = -2ia,  \vspace{2mm}  \\     \vspace{2mm}
 \left< V, \frac{\partial U}{\partial q} \right> = c,   \\
 \left< V, \frac{\partial U}{\partial r} \right> = b
 \end{array}   \eeq

 The trace identity now yields
 \beq   \begin{array}{lc}
 \frac{\delta}{\delta q} (-2ia)=\lambda^{-\gamma} \frac{\partial}{\partial \lambda}(\lambda^{\gamma} c)  \vspace{2mm}   \\    \vspace{2mm}
 \mbox{or,}  -2i \frac{\delta}{\delta q}(\sum_{m\geq 0} a_m \lambda^{-m})=\lambda^{-\gamma} \frac{\partial}{\partial \lambda}(\lambda^{\gamma} \sum_{m\geq0}c_m \lambda^{-m})  \\
 \mbox{or,} -2i \frac{\delta}{\delta q} (\sum_{m\geq0} a_m \lambda^{-m}) = \sum_{m\geq0} c_m (\gamma_{-m}) \lambda^{-m-1}
 \end{array}   \eeq
 Similarly, we have
 \beq   \begin{array}{lc}
 -2i \frac{\delta}{\delta r} (\sum_{m\geq0} a_m \lambda^{-m}) = \sum_{m\geq0} b_m (\gamma_{-m}) \lambda^{-m-1}
 \end{array}   \eeq
 Equating the coefficients of $\lambda^{-n-2}$ on both sides of $(121)$ and $(122)$, we obtain
 \beq   \begin{array}{lc}
 -2i \left( \frac{\partial}{\partial q} , \frac{\partial}{\partial r} \right) \left( a_{n+2} \right) = (\gamma_{-n-1}) (c_{n+1} , b_{n+1})
 \end{array}   \eeq
 To determine the unknown constants $\gamma$, we put $n=0$ in $(123)$ and obtain
 $$-2i \frac{\delta}{\delta q} (a_2) = (\gamma-1) c_1 $$
 Using the values of $a_2$ and $c_1$ and the definition of the variational derivative in the above equation we obtain
 $$   \begin{array}{lc}
 -2i \frac{\delta}{\delta q} (\frac{\alpha}{2} q r) = (\gamma -1) i \alpha r  \vspace{2mm}    \\     \vspace{2mm}
 \mbox{or,} -2 (\frac{\alpha}{2} r) = (\gamma -1) \alpha r
 \end{array}   $$
 which leads to $\gamma=0$     \\
 Thus we are left with
 \beq   \begin{array}{lc}
 \left( \frac{\partial}{\partial q} , \frac{\partial}{\partial r} \right) \left( 2i \frac{a_{n+2}}{n+1} \right) = (c_{n+1} , b_{n+1})  \vspace{2mm}   \\    \vspace{2mm}
 \left( \frac{\partial}{\partial q} , \frac{\partial}{\partial r} \right) H_n= (c_{n+1} , b_{n+1})
 \end{array}   \eeq
 where,
 \beq   \begin{array}{lc}
 H_n=2i\left( \frac{a_{n+2}}{n+1} \right)
 \end{array}   \eeq
 The AKNS hierarchy can now be cast in the Hamiltonian form
 \beq    \begin{array}{lc}
 u_t= \left( \matrix{q  \\  r} \right)_t = J L^n \left( \matrix{c_1  \\  b_1} \right) = J \frac{\delta H_n}{\delta u}
 \end{array}    \eeq
where $J$ and $L$ are defined previously and $H_n$ are hierarchy of commuting conserved functionals.

\subsection{Hamiltonian structure of the Kaup-Newell hierarchy}

This hierarchy is defined by
$$ \begin{array}{lc}
U=\left( \matrix{ -i \lambda^2  &  \lambda q  \\  \lambda r  &  i \lambda^2} \right)  \hspace{4mm}  \mbox{and}  \hspace{4mm}   V=\left( \matrix{ a &  b  \\  c  &  -a} \right) \end{array} $$
The hierarchy can be expressed as
\beq   \begin{array}{lc}
u_t=\left( \matrix{q  \\  r} \right)_t = \left( \matrix{ b_{(2n+1)x}  \\  c_{(2n+1)x}} \right)=J\left(  \matrix{c_{2n+1}  \\  b_{2n+1}} \right)
\end{array}   \eeq
with $J$ given by
\beq   \begin{array}{lc}
J=\left( \matrix{0   &   \partial  \\  \partial  &  0} \right)
\end{array}   \eeq
The operators $L_1$ and $L_2$ defined by
\beq   \begin{array}{lc}
L_1=\frac{1}{2} \left( \matrix{r \partial^{-1} r  &  -i+r \partial^{-1} q  \\  i + q \partial^{-1} r  &  q \partial^{-1} q} \right),
L_2=\left( \matrix{0   &   \partial  \\  \partial  &  0} \right)
\end{array}   \eeq
are such that
\beq   \begin{array}{lc}
L_1 L_2\left( \matrix{c_{2n-1}  \\  b_{2n-1}} \right) = \left( \matrix{c_{2n+1}  \\  b_{2n+1}} \right)
\end{array}   \eeq
As a check one notes that
$$  \begin{array}{lc}
\frac{1}{2} \left( \matrix{r \partial^{-1} r  &  -i+r\partial^{-1} q  \\  i+ q \partial^{-1} r  &  q \partial^{-1} q} \right) \left( \matrix{0  & \delta  \\  \delta  &  0} \right)
\left( \matrix{c_1  \\  b_1} \right)   = \left( \matrix{-i r_x + r^2 q  \\  i q_x + q^2 r} \right) = \left( \matrix{c_3  \\  b_3} \right)
\end{array}  $$
In general then
\beq   \begin{array}{lc}
(L_1 L_2)^n \left( \matrix{c_1  \\  b_1} \right) = \left( \matrix{c_{2n+1}  \\  b_{2n+1}} \right)
\end{array}   \eeq
To use the trace identity, we compute the following
\beq   \begin{array}{lc}
\left< V, \frac{\partial U}{\partial \lambda} \right> = -4 i \lambda a + r b + q c,    \vspace{2mm}      \\      \vspace{2mm}
\left< V, \frac{\partial U}{\partial q} \right> = \lambda c,    \\
\left< V, \frac{\partial U}{\partial r} \right> = \lambda b
\end{array}\eeq
The trace identity gives
\beq   \begin{array}{lc}
\frac{\delta}{\delta q} (-4 i \lambda a + r b + q c) = \lambda^{-\gamma} \frac{\partial}{\partial \lambda} (\lambda^{\gamma} \lambda c)     \vspace{2mm}    \\   \vspace{2mm}
\mbox{and} \hspace{4mm} \frac{\delta}{\delta r} (-4 i \lambda a + r b + q c) = \lambda^{-\gamma} \frac{\partial}{\partial \lambda} (\lambda^{\gamma} \lambda b)
\end{array}   \eeq
Expanding $a$, $b$ and $c$ in the first of the above expressions and equating the coefficients of $\lambda^{-(2m+1)}$ on both sides, we obtain,
$$  \frac{\delta}{\delta q} \left[ -4i a_{2m+2} + r b_{2m+1} q c_{2m+1} \right] = (\gamma -2m) c_{2m+1}  $$
Putting $m=0$ in the above equation we obtain $\gamma=0$   \\
Using the trace identity we have
\beq   \begin{array}{lc}
\left( \frac{\delta}{\delta q} , \frac{\delta}{\delta r} \right) H_m=(c_{2m+1} , b_{2m+1})
\end{array}   \eeq
with
\beq   \begin{array}{lc}
H_m=\frac{4 i a_{2m+2} - r b_{2m+1} - q c_{2m+1}}{2m}
\end{array}   \eeq
and the hierarchy given by
$$u_t=J \frac{\delta H_n}{\delta u}$$

\subsection{The Hamiltonian formalism for the general structure of the DNLS equations}

In the general case, one starts with the following spectral problem,
\beq   \begin{array}{lc}
U= \left( \matrix{ -i \lambda^2 - i \beta q r  &  \lambda q  \\  \lambda r  &  i \lambda^2 + i \beta q r} \right)
\end{array}   \eeq
as already discussed in some detail in section $4.3$ \\
Since the steps to be followed in order to obtain the Hamiltonian structure have already been explained in the preceding two subsections,for this example we just provide the outline of the procedure.    \\

The coupled system of the hierarchy of equations are given by $(99)$ and can be cast in the form
\beq   \begin{array}{lc}
\left( \matrix{q  \\  r}\right)_t= L_3 L_2\left( \matrix{c_{2n+1}  \\  b_{2n+1}} \right)
\end{array}   \eeq
where $L_3$ and $L_2$ are given by
\beq   \begin{array}{lc}
L_3= \left( \matrix{1 - 2i\beta q \partial^{-1} r  &  -2i\beta q \partial^{-1} q  \\  2i\beta r \partial^{-1} r  &  1 + 2i\beta r \partial^{-1} q} \right),
L_2=\left( \matrix{0   &   \partial + 2i\beta q r \  \\  \partial - 2i\beta q r  &  0} \right)
\end{array}   \eeq
The coefficients in the expansion of b and c are related by
\beq   \begin{array}{lc}
L_1 L_2\left( \matrix{c_{2j-1}  \\  b_{2j-1}} \right) = \left( \matrix{c_{2j+1}  \\  b_{2j+1}} \right)
\end{array}   \eeq
where
\beq   \begin{array}{lc}
L_1=\frac{1}{2} \left( \matrix{r \partial^{-1} r  &  -i+r \partial^{-1} q  \\  i + q \partial^{-1} r  &  q \partial^{-1} q} \right),
\end{array}   \eeq
and $L_2$ is already defined above.
Here $L_1$ and $L_2$ are skew-symmetric operators i.e.
$$L_1^* = -L_1$$ and  $$L_2^* = -L_2$$
Extending $(139)$ we can write
\beq   \begin{array}{lc}
(L_1 L_2)^n \left( \matrix{c_1  \\  b_1} \right) = \left( \matrix{c_{2n+1}  \\  b_{2n+1}} \right)
\end{array}   \eeq
where $$c_1 = 2r \hspace{4mm}  \mbox{and} \hspace{4mm}  b_1=2q$$.
Let us now introduce the function
\beq   \begin{array}{lc}
P_{2j+1} = \left( \matrix{ c_{2j+1} - 2i \beta r a_{2j}  \\  b_{2j+1} - 2i \beta q a_{2j}} \right)
\end{array}   \eeq
It can be shown that,
\beq   \begin{array}{lc}
L_3^* P_{2j+1} = \left(  \matrix{c_{2j+1}  \\  b_{2j+1}} \right),  \hspace{4mm}  j \geq 0
\end{array}   \eeq
where we need to use the relation
$$a_{2jx} = q c_{2j+1} - r b_{2j+1}$$
given by equation $(88)$ and $L_3^*$ denotes the conjugate of $L_3$ .   \\
The hierarchy of equations can now be written as
\beq   \begin{array}{lc}
u_t=\left(  \matrix{ q  \\  r} \right)_t = L_3 L_2 \left( \matrix{c_{2n+1}  \\  b_{2n=1}} \right) = L_3 L_2 L_3^* P_{2n+1}    \vspace{2mm}     \\     \vspace{2mm}
\mbox{or,}  \hspace{4mm}  u_t=J P_{2n+1}   \hspace{4mm}  \mbox{where}  \\
J = L_3 L_2 L_3^*
\end{array}   \eeq
It may further be shown that the operators $J L^k (k=0,1,2, ... m)$ are skew-symmetric.
For the trace identity, we compute
\beq   \begin{array}{lc}
\left< V, \frac{ \partial U}{ \partial \lambda} \right> = -4 i \lambda a + r b + q c      \vspace{2mm}     \\     \vspace{2mm}
\left< V, \frac{ \partial U}{ \partial q} \right> = c \lambda - 2 i \beta r a   \\
\left< V, \frac{ \partial U}{ \partial r} \right> = b \lambda - 2 i \beta q a
\end{array}   \eeq
The trace identity gives,
\beq   \begin{array}{lc}
\frac{\delta}{\delta u} ( -4 i \lambda + r b + q c) = \lambda^{- \gamma} \frac{\partial}{\partial \lambda} \left[ \lambda^{\gamma} (\lambda c - 2  i\beta r a, \lambda b - 2 i \beta q a) \right]
\end{array}   \eeq
Expanding a, b and c in negative powers of $\lambda$ and equating coefficients of $\lambda^{-(2n+1)}$, we arrive at the relation
\beq   \begin{array}{lc}
\frac{\delta}{\delta u} (-4 i a_{2n+2} + r b_{2n+1} + q c_{2n+1}) = (\gamma -2n) (c_{2n+1} - 2 i \beta r a_{2n}, b_{2n+1} - 2 i \beta q a_{2n})
\end{array}  \eeq
Putting $n=0$, we find $\gamma=0$.  \\
Therefore,
\beq   \begin{array}{lc}
\frac{\delta}{\delta u} \left[ \frac{4 i a_{2n+2} - r b_{2n+1} q c_{2n+1}}{2n} \right]     \vspace{2mm}      \\      \vspace{2mm}
= (c_{2n+1} - 2 i \beta r a_{2n} , b_{2n+1} - 2 i \beta q a_{2n})   \\
\mbox{or,}  P_{2n+1} = \frac{\delta H_n}{\delta u}
\end{array}   \eeq
where
\beq   \begin{array}{lc}
H_0=2 q r, \hspace{4mm}  H_n=\frac{4 i a_{2n+2} - r b_{2n+1} - q c_{2n+1}}{2n}, \hspace{4mm}  n\geq 1
\end{array}   \eeq
This means that the hierarchy can be expressed as
$$ u_t=J \frac{\delta H_n}{\delta u}$$
In view of the theory outlined previously, the complete integrability of the system is established.  \\
It is worthwhile to note that the hierarchy of the evolution equations in all the cases discussed above can be put in the form
\beq   \begin{array}{lc}
u_t=J \frac{\delta H_n}{\delta u} = J L \frac{\delta H_{n-1}}{\delta u} = ...... = J L^n \frac{\delta H_0}{\delta u},  \hspace{4mm}  n=1,2,3 ....
\end{array}   \eeq
in the light of the forgoing discussion where $J$ and $L$ are already defined along with their properties.   \\
Equation  $(150)$ emphasizes the multi-Hamiltonian formulation of the hierarchy of non-linear evolution equations in all the different cases discussed above.

\section{The Non-holonomic Deformation of Integrable Systems}

It would be pertinent to explain what exactly is meant by the Non-holonomic deformation (NHD) of integrable systems. Perturbation generally disturbs the integrability of a system. However, when we consider NHD of an integrable system, the system gets perturbed with a deforming function in such a way that under suitable differential constraints on the perturbing function, the system maintains its integrability. The constraints are furnished in the form of differential relations and they turn out to be equivalent to a non-holonomic constraint.   \\
To construct these non-holonomic deformations, one starts with a lax pair, keeping the space part $U(\lambda)$ unchanged but modifying the temporal component $V(\lambda)$. This implies that the scattering problem remain unchanged, but the time evolution of the spectral data becomes different in the perturbed models. Corresponding to these deformed systems, it is possible to generate some kind of two-fold integrable hierarchy. One method is to keep the perturbed equations the same but increase the order of the differential constraints in a recursive manner, thus generating a new integrable hierarchy for the deformed system. Alternatively, the constraint may be kept fixed at its lowest level, but the order of the original equation may be increased in the usual way, thereby leading to new hierarchies of integrable systems.   \\

\subsection{Non-Holonomic Deformation (NHD) of the Non-linear Schr\"odinger Equation (NLSE)}

The spatial and temporal components of the Lax pair of the NLS equation are given by,

\beq   \begin{array}{lc}
U=-i \lambda \sigma_3 +q \sigma_+ + r \sigma_-
\end{array}   \eeq

\beq   \begin{array}{lc}
V_{original}=-i \lambda^2 \sigma_3 + \lambda(q \sigma_+ + r\sigma_-) + {-(\frac{i}{2})qr\sigma_3 + (\frac{i}{2}q_x)\sigma_+ - (\frac{i}{2}r_x)\sigma_-}
\end{array}   \eeq
To obtain the deformation of the NLS equation, let us inroduce
\beq   \begin{array}{lc}
V_{deformed}=\frac{i}{2}\lambda^{-1} G^{(1)}
\end{array}   \eeq
where
\beq   \begin{array}{lc}
G^{(1)}=a\sigma_3+g_1\sigma_++g_2\sigma_-
\end{array}   \eeq
So that the time part of the Lax pair takes the form
\beq   \begin{array}{lc}
\tilde{V}=V_{original} + V_{deformed}
\end{array}   \eeq   \\[2mm]

Let us now impose the zero curvature or the flatness condition
$$U_t-\tilde{V}_x+[U,\tilde{V}]=0$$ with $U$ and $V$ as in $(151)$ and $(155)$ respectively.

We observe from the zero curvature condition that while the positive powers of $\lambda$ are trivially satisfied,
the zeroth power (or the $\lambda$ free term) leads to the perturbed dynamical systems (equations), while the negative
powers of $\lambda$ give rise to the differential constraints.  \\

For example, the deformed pair of the NLS equations are given by,
\beq   \begin{array}{lc}
q_t-\frac{i}{2}q_{xx}+iq^2r=-g_1
\end{array}   \eeq

\beq   \begin{array}{lc}
r_t+\frac{i}{2}r_{xx}-iqr^2=g_2
\end{array}   \eeq

Considering the $\lambda^{-1}$ terms and equating the coefficients of the generators $\sigma_3$, $\sigma_+$, $\sigma_-$ successively, we obtain the following individual constraint conditions on the functions $a$, $g_1$ and $g_2$
\beq   \begin{array}{lc}
a_x = q g_2 - r g_1
\end{array}   \eeq
\beq   \begin{array}{lc}
g_{1x} + 2aq = 0
\end{array}   \eeq
\beq   \begin{array}{lc}
g_{2x} - 2ar = 0
\end{array}   \eeq

The foregoing equations can be shown to give rise to the differential constraint

\beq   \begin{array}{lc}
\hat{L}(g_1, g_2) = r g_{1xx} + q_x g_{2x} + 2qr (q g_2 - r g_1) = 0
\end{array}   \eeq

Eliminating the deforming functions $g_1$ and $g_2$, we can derive a new higher order equation as

\beq   \begin{array}{lc}
-r (q_t - \frac{i}{2} q_{xx} + i q^2 r)_{xx} + q_x (r_t + \frac{i}{2} r_{xx} - i q r^2)_x \\ +  2qr [q (r_t + \frac{i}{2} r_{xx} - i q r^2) + r(q_t - \frac{i}{2} q_{xx} + iq^2 r)] = 0
\end{array}   \eeq

We can now consider a double deformation of the NLS equation by taking

\beq   \begin{array}{lc}
V_{deformed}(\lambda) = \frac{i}{2} (\lambda^{-1} G^{(1)} + \lambda^{-2} G^{(2)})
\end{array}   \eeq

where the function $G^{(2)}$ is given by

\beq   \begin{array}{lc}
G^{(2)} = b\sigma_3 + f_1 \sigma_+ + f_2 \sigma_-
\end{array}   \eeq

and $G^{(1)}$ is already defined in equation $(154)$.

The zero-curvature condition is now applied with $U$ as before but $V_{deformed}$ as defined in $(163)$. The following results arise:

(i) No change occurs in the deformed NLS equations.\\
(ii) Picking up the terms in $\lambda^{-1}$ and equating the coefficients of the generators $\sigma_3$, $\sigma_+$ and $\sigma_-$ successively, we are led to the following individual constraints

\beq   \begin{array}{lc}
a_x = q g_2 - r g_1
\end{array}   \eeq

\beq   \begin{array}{lc}
g_{1x} + 2 i f_1 + 2 a q = 0
\end{array}   \eeq

\beq   \begin{array}{lc}
g_{2x} - 2 i f_2 - 2 a r = 0
\end{array}   \eeq
The preceding set of equations finally lead to the following differential constraint

\beq   \begin{array}{lc}
\hat{L(g_1,g_2)}+2i(r f_{1x}-q_x f_2)=0
\end{array}   \eeq
with
\beq   \begin{array}{lc}
\hat{L}(g_1,g_2)=r g_{1xx} + g_{2x} q_x + 2qr(q g_2-r g_1)
\end{array}   \eeq

(iii) The terms in $\lambda^{-2}$ give rise to a second constraint

\beq   \begin{array}{lc}
\hat{L (f_1, f_2)} = 0
\end{array}   \eeq

where the functional form of the above expression is already given by $(169)$ while $f_1$, $f_2$ make up the argument in $(170)$.

Thus, this is an example where the perturbed equations are kept the same, but the order of the differential constraint is increased recursively, thereby creating a new integrable hierarchy for the NLS equation.

\subsection{NHD of coupled KdV type NLSE}

For the coupled KdV type NLSE, the space and time components of the Lax pair are given by,

\beq   \begin{array}{lc}
U=-i\lambda \sigma_3 + q \sigma_+ + r \sigma_-   \hspace{4mm}  \mbox{and}    \\
V_{original}=-i\lambda^3 \sigma_3 + \lambda^2(q \sigma_+ + r \sigma_-) + \lambda[(-\frac{i}{2}q r)\sigma_3 + \frac{i}{2}q_x \sigma_+ -\frac{i}{2} r_x \sigma_-] \\
+[\frac{1}{4}(r q_x - q r_x)\sigma_3 + (\frac{1}{2}q^2 r - \frac{1}{4}q_{xx})\sigma_+ + (\frac{1}{2}q r^2 - \frac{1}{4}r_{xx})\sigma_-]
\end{array}   \eeq

Note that $V_{original}$ now includes a term in $\lambda^3$ as compared to $\lambda^2$ in the previous example of the NLS equation. This would lead to a higher order dispersion term.

Take $V_{deformed}$  = $\frac{i}{2}\lambda^{-1}G^{(1)}$

therefore, $\tilde{V}$ = $V_{original} + V_{deformed}$.

Using the zero-curvature condition, we arrive at the following deformed equations:

\beq   \begin{array}{lc}
q_t + \frac{1}{4} q_{xxx} - \frac{3}{2} q q_x r = - g_1
\end{array}   \eeq

and

\beq   \begin{array}{lc}
r_t + \frac{1}{4} r_{xxx} - \frac{3}{2} r r_x q = g_2
\end{array}   \eeq

along with the differential constraint

\beq   \begin{array}{lc}
\hat{L} (g_1, g_2) = 0
\end{array}   \eeq

In this example, the constraint is held fixed at its lowest level, but the order of the NLS equation is increased (terms enter with higher order dispersion) and thus a new integrable hierarchy can be formed.

The generalized NLSE is actually a combination of the ordinary NLSE and the coupled KdV type NLSE. NHD of such a system can be carried out in the manner already outlined previously.

\subsection{NHD of Derivative NLS equation (DNLS) : Kaup-Newell (KN) system}

In this case, the Lax pair are given by

\beq   \begin{array}{lc}
U = -i \lambda^2 \sigma_3 + \lambda (q \sigma_+ + r \sigma_-)
\end{array}   \eeq

\beq   \begin{array}{lc}
V_{original} = - i \lambda^4 \sigma_3 + \lambda^3 (q \sigma_+ + r \sigma_-) - \lambda^2 q r (\frac{i}{2})\sigma_3 + \\
\lambda [(\frac{1}{2} q^2 r + \frac{i}{2} q_x)\sigma_+ + (\frac{1}{2} q r^2 - \frac{i}{2} r_x) \sigma_-]
\end{array}   \eeq

The modified temporal component of the Lax pair is given as

$\tilde{V}$ = $V_{original} + V_{deformed}$

where

\beq   \begin{array}{lc}
V_{deformed} = i (G^{(0)} + \lambda^{-1} G^{(1)} + \lambda^{-2} G^{(2)})
\end{array}   \eeq

and

\beq   \begin{array}{lc}
G^{(0)} = w \sigma_3 + m_1 \sigma_+ + m_2 \sigma_-
\end{array}   \eeq
\beq   \begin{array}{lc}
G^{(1)} = a \sigma_3 + g_1 \sigma_+ + g_2 \sigma_-
\end{array}   \eeq
\beq   \begin{array}{lc}
G^{(2)} = b \sigma_3 + f_1 \sigma_+ + f_2 \sigma_-
\end{array}   \eeq

Using the zero-curvature relation, we obtain the following deformed DNLS equations:

\beq   \begin{array}{lc}
q_t - \frac{i}{2} q_{xx} - \frac{1}{2} (q^2 r)_x + 2 g_1 - 2 i q w = 0
\end{array}   \eeq
\beq   \begin{array}{lc}
r_t + \frac{i}{2} r_{xx} - \frac{1}{2}(q r^2)_x - 2 g_2 + 2 i r w = 0
\end{array}   \eeq

Further, we obtain the following conditions on the different components of the deforming functions $G^{(i)}$:
$m_1 = 0$, $m_2 = 0$, $a = 0$, $f_1 = 0$, $f_2 = 0$ and $b_x = 0$ which implies that $b = b(t)$ only.

We are, therefore, left with the following deforming functions:
\beq   \begin{array}{lc}
G^{(0)} = w (x, t) \sigma_3\\

G^{(1)} = g_1 (x, t) \sigma_+ + g_2 (x, t) \sigma_-\\

G^{(2)} = b (t) \sigma_3
\end{array}   \eeq

Moreover, the following constraints are obtained:

\beq   \begin{array}{lc}
g_{1x} = - 2 q (x, t) b (t)\\

g_{2x} =  2 r (x, t) b (t)\\

w_x = q g_2 - r g_1
\end{array}   \eeq

It is possible to obtain new non-linear integrable equations by resolving the constraint relations and expressing all the perturbing functions through the basic field variables. To this end, we put

\beq   \begin{array}{lc}
q = u_x\\
r = v_x
\end{array}   \eeq

where $u = u (x, t)$ and $v = v (x, t)$

Equation $(185)$ used in equation $(184)$ allows us to express $g_1$, $g_2$ and w in terms of $b (t)$, u and v only as follows :

\beq   \begin{array}{lc}
g_1 = -2 b (t) u\\
g_2 = 2 b(t) v\\
w = 2 b(t) u v + K (t)
\end{array}   \eeq
where K is again a function of t only. \\

Eliminating $g_1$, $g_2$ and w from  equations $(181)$ and $(182)$, we can rewrite the coupled perturbed $(deformed)$ DNLS equations in the following form:

\beq   \begin{array}{lc}
u_{xt} - \frac{i}{2} u_{xxx} - \frac{1}{2} (u_x^2 v_x)_x - 4 u b (t) - 2 i u_x (2 b (t) u v + K (t)) = 0
\end{array}   \eeq
\beq   \begin{array}{lc}
v_{xt} + \frac{i}{2} v_{xxx} - \frac{1}{2} (u_x v_x^2)_x - 4 v b (t) + 2 i v_x (2 b (t) u v + K (t)) = 0
\end{array}   \eeq

These are coupled evolution equations which are non-autonomous with arbitrary time-dependent coefficients $b (t)$ and $K ( t)$. Clearly, no more constraints are left at this stage. Equations $(187)$ and $(188)$ generalize the coupled system of Lenells-Fokas equations $[38]$, $[39]$ by including a non-linear derivative term as well as a higher order dispersion term.

\subsection{NHD of Chen-Lee-Liu (CLL) system}

The Lax pair for the CLL equations is given by,
\beq   \begin{array}{lc}
U = \lambda^2 \left(\matrix{    -i & 0 \\ 0 & i  }    \right) + \lambda \left( \matrix { 0 & q \\ r & 0} \right) + \left (\matrix { 0 & 0 \\ 0 & \frac{i}{2} q r} \right)
\end{array}   \eeq

\beq   \begin{array}{lc}
  V = 2 \lambda^4 \left(\matrix {   -i  &  0  \\   0  &  i}  \right)  +  2 \lambda^3 \left(\matrix {   0  &  q  \\   r  &  0}  \right) +  \lambda^2 q r \left(\matrix {   -i  &  0  \\   0  &  i}  \right) +  \lambda \left(\matrix {   0  &  i q_x + \frac{1}{2} q^2 r  \\   - i r_x + \frac{1}{2} q r^2  &  0}  \right)\\ + \left(\matrix {   0  &  0  \\   0  &  - \frac{1}{2} (r q_x - r_x q) + \frac{i}{4} r^2 q^2}  \right)
\end{array}   \eeq

We take
\beq   \begin{array}{lc}
V_{deformed} = i (G^{(0)} + \lambda^{-1} G^{(1)} + \lambda^{-2} G^{(2)})
\end{array}   \eeq

Here $G^{(0)} = w \sigma_3$, $G^{(1)} = g_1 \sigma_+ + g_2 \sigma_-$, $G^{(2)} = b \sigma_3$

where we have taken the cue from the discussion in the previous section in choosing the form of the matrices $G^{(0)}$, $G^{(1)}$ and $G^{(2)}$.\\

Taking  $\tilde{V}$ = $V + V_{deformed}$, and imposing the zero-curvature condition, we are led to the following deformed CLL equations:

\beq   \begin{array}{lc}
q_t = i q_{xx} + q q_x r - 2 g_1 + 2 i q w
\end{array}   \eeq

\beq   \begin{array}{lc}
r_t = - i-r_{xx} + r r_x q + 2 g_2 - 2 i r w
\end{array}   \eeq

The following differential constraints are also obtained:

\beq   \begin{array}{lc}
i w_x = q g_2 - r g_1
\end{array}   \eeq
\beq   \begin{array}{lc}
g_{1x} + 2 q b + \frac{i}{2} q r g_1 = 0
\end{array}   \eeq
\beq   \begin{array}{lc}
g_{2x} - 2 r b - \frac{i}{2} q r g_2 = 0
\end{array}   \eeq

We also get $b_x = 0$ which implies that b is a function of t only. \\
However, it may be noted that it is not possible in this case to resolve the constraints and express the perturbing functions through the basic field variables by re-defining these variables. This is due to the presence of a non-linear term in $(195)$ and $(196)$. \\

It may be mentioned in passing that the NHD of the Kundu-Eckhaus equation can be worked out in an exactly similar manner. However, we are not reproducing the details of that calculation here.

\subsection{NHD of the hierarchy of equations in the AKNS system}

The hierarchy of dynamical equations in the AKNS system is given by
\beq   \begin{array}{lc}
q_t = - 2 i b_{n + 1}\\
r_t = 2 i c_{n + 1}
\end{array}   \eeq
 
Successive equations can be generated by putting n = 1, 2, 3 etc. \\
Since the time part of the Lax pair is given by
\beq   \begin{array}{lc}
V^{(n)} = (\lambda^n V)_+ + \triangle_n = (\lambda^n V)_+ = \sum_{m=0}^{n}\lambda^{n-m} (a_m \sigma_3 + b_m \sigma_+ + c_m \sigma_-)
\end{array}   \eeq
we introduce the non-holonomic deformation by taking
\beq   \begin{array}{lc}
V_{deformed} = \frac{i}{2} (\lambda^{-1} G^{(1)} + \lambda^{-2} G^{(2)} + \lambda^{-3} G^{(3)} + ..... )
\end{array}   \eeq

where $$G^{(1)} = a_1 \sigma_3 + g_1 \sigma_+ + g_2 \sigma_-$$\\
with similar expressions for $G^{(2)}$, $G^{(3)}$ etc.

Taking $$V_{final} = V^{(n)} + V_{deformed}$$ and using the zero-curvature condition with $V_{final}$ as the (new) time part of the Lax pair, we get the following deformed equations:
\beq   \begin{array}{lc}
q_t = - 2 i b_{n + 1} - g_1\\
r_t = 2 i c_{n + 1} + g_2
\end{array}   \eeq
along with the constraint conditions given as a hierarchy of recursive relations as follows:
\beq   \begin{array}{lc}
i G^{(1)}_x = [\sigma_3, G^{(2)}] + i [q \sigma_+ + r \sigma_-, G^{(1)}]
\end{array}   \eeq
\beq   \begin{array}{lc}
i G^{(2)}_x = [\sigma_3, G^{(3)}] + i [q \sigma_+ + r \sigma_-, G^{(2)}]
\end{array}   \eeq

and so on.

\subsection{NHD of the hierarchy of equations in the DNLS system (Kaup-Newell hierarchy)}

In this section, we show how non-holonomic deformation may be applied to the equations of the DNLS hierarchy (KN system) obtained by using the Tu methodology. \\
The space and time components of the Lax pair are given as follows:
\beq   \begin{array}{lc}
U = - i \lambda^2 \sigma_3 + \lambda q \sigma_+ + \lambda r \sigma_-
\end{array}   \eeq
\beq   \begin{array}{lc}
\tilde{V^{(n)}} = \sum_{j=0}^{n}(a_{2j} \lambda^{2(n - j)+ 2} \sigma_3 + b_{2j + 1} \lambda^{2(n-j) + 1} \sigma_+ + c_{2j + 1} \lambda^{2(n - j) + 1} \sigma_-)
\end{array}   \eeq

On using the zero curvature equation, we get the hierarchy of equations for the Kaup-Newell system:

\beq   \begin{array}{lc}
q_t = b_{(2n + 1)x}\\
r_t = c_{(2n + 1)x}
\end{array}   \eeq

To carry out the non-holonomic deformation, we take
\beq   \begin{array}{lc}
V_{deformed} = \frac{i}{2} (G^{(0)} + \lambda^{-1} G^{(1)} + \lambda^{-2} G^{(2)} + \lambda^{-3} G^{(3)} + ....... )
\end{array}   \eeq

so that the time part of the Lax pair becomes
\beq   \begin{array}{lc}
V_{final} = \tilde{V^{(n)}} + V_{deformed}
\end{array}   \eeq

Now applying the zero curvature condition again with U and $V_{final}$ as the Lax pair, we get the deformed equations of the Kaup-Newell hierarchy as:
\beq   \begin{array}{lc}
q_t = b_{(2 n + 1) x} - g_1 + i q a_0\\
r_t = c_{(2n + 1)x} + g_2 - i r a_0
\end{array}   \eeq

where it has been deduced that
\beq   \begin{array}{lc}
G^{(0)} = a_0 \sigma_3
\end{array}   \eeq

and $G^{(1)}$ is taken to be
\beq   \begin{array}{lc}
G^{(1)} = a_1 \sigma_3 + g_1 \sigma_+ + g_2 \sigma_-
\end{array}   \eeq

The differential constraints are given recursively by a series of equations as follows:
\beq   \begin{array}{lc}
G^{(0)}_x = - i [\sigma_3, G^{(2)}] + [q \sigma_+ + r \sigma_-, G^{(1)}]
\end{array}   \eeq
\beq   \begin{array}{lc}
G^{(1)}_x = - i [\sigma_3, G^{(3)}] + [q \sigma_+ + r \sigma_-, G^{(2)}]
\end{array}   \eeq
etc.

\bigskip

\section{Discussion}
The family of Non-linear Schrodinger equations have been studied exhaustively by using two different techniques viz. the AKS framework and the Tu methodology. In this
section we try to explore the connection between these two formalisms.
The construction of an integrable dynamical system is accomplished by using the zero-curvature equation. This means we need to obtain both the spatial and temporal components
of the Lax pair. Applying the zero-curvature condition on the Lax pair would lead us to the integrable system in one space and one temporal dimension.
Both the methods used in this work start by identifying a properly chosen spectral problem through the space part of the Lax pair. In the AKS method, this object, i.e.
the orbit can be constructed by suitable co-adjoint action of the Lie group acting on an element of the Lie algebra. Thus the underlying geometry of the integrable
model gets emphasized in the AKS method.
In order to obtain the temporal component of the Lax pair in an Infinite Dimensional Lie Algebra, in the AKS method, one expands this temporal component in powers
of the spectral parameter and take a suitable projection on a particular subalgebra of the IDLA. Once the temporal component is determined by this technique, application
of zero-curvature leads to the desired equation for the dynamical system. The commuting flows of the AKS hierarchy can also be obtained in this framework.
In the Tu methodology also, we expand the time component V in negative powers of the spectral parameter and use it in the stationary zero-curvature equation
to obtain the different elements of V  in a recursive manner. After this, the expansion in negative powers of $\lambda$ is multiplied by a suitable positive degree of $\lambda$ and the
projection taken on non-negative powers of $\lambda$. Zero curvature is subsequently applied, with a suitable constraint to obtain the hierarchy of non-linear evolution equations.
Thus, in the Tu method we need to choose the spectral problem judiciously and the hierarchy results naturally when the sequence of steps outlined above is applied. While both
the AKS and Tu methods rely on an expansion of the temporal component of the Lax pair in powers of $\lambda$, the AKS method definitely stresses on the geometry
underlying the construction of the integrable system, whereas the Tu method is more algebraic in spirit.
Use of the trace identity to determine the Hamiltonian and then application of the operators J and L to set up the complete Hamiltonian structure is another remarkable
feature of the Tu method. The AKS scheme also endows an Integrable System with a Hamiltonian structure. But the trace identity method is definitely a more convenient
tool to set up the Hamiltonian or rather the multi-Hamiltonian structure of the Integrable System. Associated results then guarantee the complete integrability of the system.
One feels that using the AKS theorem and the Tu methodology in tandem will help us unearth rich results in the domain of non-linear Integrable Systems.In particular,
applying both techniques to the same class or family of problems will help us in understanding the problem in finer detail.
We have also carried out a detailed analysis covering Non-holonomic deformation of different equations of the NLS family. In particular, NHD has been applied to the hierarchy of equations (AKNS and DNLS-Kaup-Newell systems) obtained by using the Tu methodology. It may be mentioned that the structure of the DNLS system may be made more general and NHD may be applied on the resulting hierarchy. The topics covered above and related problems will be the subject of our future investigation. 

\section*{Acknowledgements}
The work owes a lot to our past discussions, correspondences and collaboration with Walter Oevel, Franco Magri, 
Marco Pedroni, Darryl Holm, Tudor Ratiu, Alfred Ramani, Victor Kac, Peter Olver, 
Sarbarish Chakravarty, Wen Xiu Ma, Asesh Roy Chowdhury and 
Sudipto Roy Choudhury. Finally we are also grateful to Allan Fordy for various references.

\end{document}